\documentclass[a4paper,11pt]{article}
\pdfoutput=1 

\usepackage{jheppub} 

\usepackage[T1]{fontenc} 
\usepackage{gensymb}

\usepackage{bbm,latexsym,amssymb}
\usepackage{color}

\def\be{\begin{equation}}
\def\ee{\end{equation}}
\def\ba{\begin{eqnarray}}
\def\ea{\end{eqnarray}}
\def\ra{\rightarrow}

\def\m#1{m_{#1}}

\def\hpm{H^\pm}

\def\mhpm{\m{\hpm}}
\def\ltap{\;\centeron{\raise.35ex\hbox{$<$}}{\lower.65ex\hbox{$\sim$}}\;}
\def\gtap{\;\centeron{\raise.35ex\hbox{$>$}}{\lower.65ex\hbox{$\sim$}}\;}

\newcommand{\bea}{\begin{eqnarray}}
\newcommand{\eea}{\end{eqnarray}}

\begin{document}

\title{Large pseudoscalar Yukawa couplings in the complex 2HDM}

\author[a]{Duarte Fontes,}
\author[a]{Jorge C. Rom\~{a}o,}
\author[b,c]{Rui Santos,}
\author[a,1]{and Jo\~{a}o P. Silva}


\affiliation[a]{
CFTP, Departamento de F\'{\i}sica, Instituto Superior T\'{e}cnico, Universidade de Lisboa,
Avenida Rovisco Pais 1, 1049 Lisboa, Portugal}
\affiliation[b]{Instituto Superior de Engenharia de Lisboa - ISEL,
	1959-007 Lisboa, Portugal}
\affiliation[c]{Centro de F\'{\i}sica Te\'{o}rica e Computacional,
    Faculdade de Ci\^{e}ncias,
    Universidade de Lisboa,
   Campo Grande, Edif\'{\i}cio C8
   P-1749-016 Lisboa, Portugal}


\emailAdd{duartefontes@tecnico.ulisboa.pt}
\emailAdd{jorge.romao@tecnico.ulisboa.pt}
\emailAdd{rasantos@fc.ul.pt}
\emailAdd{jpsilva@cftp.ist.utl.pt}

\date{\today}

\abstract{
We start by presenting the current status of a complex flavour conserving two-Higgs doublet model.
We will focus on some very interesting scenarios where unexpectedly the light Higgs couplings to
leptons and to b-quarks can have a large pseudoscalar component with a vanishing scalar component.
Predictions for the allowed parameter space at end of the next run with a total collected luminosity
of $300 \, fb^{-1}$ and $3000 \, fb^{-1}$ are also discussed. These scenarios are not excluded by present
data and most probably will survive the next LHC run. However, a measurement of the mixing angle
$\phi_\tau$, between the scalar and pseudoscalar component of the 125 GeV Higgs, in the decay
$h \to \tau^+ \tau^-$ will be able to probe many of these scenarios, even with low luminosity.
Similarly, a measurement of $\phi_t$ in the vertex $\bar t t h$ could help to
constrain the low
$\tan \beta$ region in the Type I model.
}

\preprint{CFTP/15-003}

\maketitle

\section{Introduction}
\label{sec:intro}

The discovery of the Higgs boson at the Large Hadron Collider (LHC) by the ATLAS~\cite{ATLASHiggs}
and CMS~\cite{CMSHiggs} collaborations has ignited a very large number of studies in the context
of multi-Higgs models.
It is now clear that some features of the Higgs couplings to fermions and gauge bosons have to be well within the Standard
Model (SM) predictions. Also, even if other heavy scalars are far from being experimentally excluded,
there is still no hint of scalar particles other than the 125 GeV one.
However, even if no large deviations from the SM were found, many of its extension
are still in agreement with all experiment data. Many models provide interesting
scenarios that can be probed at the next LHC run while contributing to solve some of
the outstanding problems in particle physics. Such is the case of the complex
two-Higgs double model (C2HDM). The 2HDM
was first proposed by T.~D.~Lee~\cite{Lee:1973iz} as an attempt to understand the matter-antimatter
asymmetry of the universe (the 2HDM is described in detail in~\cite{hhg, ourreview}).

The 2HDM is a simple extension of the SM where the potential is still invariant under
$SU(2) \times U(1)$ but is now built with two complex scalar doublets. The complex
two-Higgs doublet model is the version of the model that allows for CP-violation
in the potential, providing therefore an extra source of CP-violation to the theory.
The existing experimental data and in particular the one recently analysed at the LHC
has been used in several studies with the goal of constraining the parameter space of the
C2HDM~\cite{Barroso:2012wz, Inoue:2014nva,Cheung:2014oaa, Fontes:2014xva} or just
the Yukawa couplings~\cite{Brod:2013cka}.

The main purpose of this work is to analyse scenarios in the C2HDM that deviate
from the SM predictions, while being in agreement with all available experimental
and theoretical constraints. These are scenarios where the scalar component of the
Higgs coupling to leptons or to b-quarks vanishes. The respective pseudoscalar
component has to be non-zero which does not necessarily imply a very large CP-violating parameter.
Even if the scalar component is not exactly zero, there are still Yukawa couplings
where the pseudoscalar component can be much larger than the corresponding scalar
component.

We will start by discussing the status of the C2HDM. Presently the
processes $pp \to h \to WW (ZZ)$, $pp \to h \to \gamma \gamma$ and
$pp \to h \to \tau^+ \tau^-$ are measured with an accuracy of about $20$\%.
On the other hand $pp \to V (h \to b \bar b)$ has been measured at the Tevatron and at the LHC
with an accuracy of about $50$\%~\cite{cms:bb, Tuchming:2014fza} while
for $pp \to h \to Z \gamma$ an upper bound of the  order of ten times the
SM expectation at the $95$\% confidence level was found~\cite{atlas:Zph, cms:Zph}.
In order to understand how the model will perform at the end of the next
LHC run we use the expected precisions on the signal strengths of different Higgs
decay modes by the ATLAS~\cite{ATLASpred} and CMS~\cite{CMSpred}
collaborations (see also~\cite{Dawson:2013bba}) for $\sqrt{s}=14$ TeV and for 300 and 3000 $fb^{-1}$
of integrated luminosities. As previously shown in~\cite{Fontes:2014xva}, the final states
$VV$, $\gamma \gamma$ and $\tau^+ \tau^-$ are enough to reproduce quantitatively the effect
of all possible final states in the Higgs decay. Therefore, taking into account the predicted precision
for the signal strength, we will consider the situations where, at $13$ TeV, the rates are
measured within either $10$\% or $5$\% of the SM prediction. We should note that no difference can
be seen in the plots when the energy is changed from $13$ to $14$ TeV as discussed in~\cite{Fontes:2014xva}.

This paper is organized as follows. In Section~\ref{sec:model}, we describe
the complex 2HDM and the constraints imposed by theoretical and phenomenological
considerations including the most recent LHC data. In Section~\ref{status}
we discuss the present status of the model and in Section~\ref{zero} we
discuss the scenarios where the pure scalar component of the Yukawa coupling
is allowed to vanish. Our conclusions are presented in Section~\ref{sec:conc}.


\section{The complex 2HDM}
\label{sec:model}

The complex 2HDM was recently reviewed in great detail in~\cite{Fontes:2014xva}
(see also ~\cite{Ginzburg:2002wt, Khater:2003wq, ElKaffas:2007rq, ElKaffas:2006nt,
Grzadkowski:2009iz, Arhrib:2010ju, Barroso:2012wz}).
Therefore, in this section we will just briefly describe the main features of the
complex two two-Higgs doublet with a softly broken $Z_2$ symmetry $\phi_1 \ra \phi_1, \phi_2 \ra -\phi_2$
whose scalar potential we write as~\cite{ourreview}
\ba
V_H
&=&
m_{11}^2 |\phi_1|^2
+ m_{22}^2 |\phi_2|^2
- m_{12}^2\, \phi_1^\dagger \phi_2
- (m_{12}^2)^\ast\, \phi_2^\dagger \phi_1
\nonumber\\*[2mm]
&&
+\, \frac{\lambda_1}{2} |\phi_1|^4
+ \frac{\lambda_2}{2} |\phi_2|^4
+ \lambda_3 |\phi_1|^2 |\phi_2|^2
+ \lambda_4\, (\phi_1^\dagger \phi_2)\, (\phi_2^\dagger \phi_1)
\nonumber\\*[2mm]
&&
+\, \frac{\lambda_5}{2} (\phi_1^\dagger \phi_2)^2
+ \frac{\lambda_5^\ast}{2} (\phi_2^\dagger \phi_1)^2.
\label{VH}
\ea
All couplings except  $m_{12}^2$ and $\lambda_5$ are real due to the hermiticity
of the potential. The complex 2HDM model as first defined in~\cite{Ginzburg:2002wt},
is obtained by forcing $\textrm{arg}(\lambda_5) \neq 2\, \textrm{arg}(m_{12}^2)$
in which case the two phases cannot be removed simultaneously.
From now on we will refer to this model as C2HDM.

We choose a basis where the vacuum expectation values (vevs) are real. Whenever we refer to the
CP-conserving 2HDM, not only the vevs, but also $m_{12}^2$ and $\lambda_5$ are taken real.
Therefore, 2HDM refers to a softly broken $Z_2$ symmetric model where all parameters of the
potential and the vevs are real. Writing the scalar doublets as
\be
\phi_1 =
\left(
\begin{array}{c}
\varphi_1^+\\
\tfrac{1}{\sqrt{2}} (v_1 + \eta_1 + i \chi_1)
\end{array}
\right),
\hspace{5ex}
\phi_2 =
\left(
\begin{array}{c}
\varphi_2^+\\
\tfrac{1}{\sqrt{2}} (v_2 + \eta_2 + i \chi_2)
\end{array}
\right),
\ee
with $v = \sqrt{v_1^2 + v_2^2} = (\sqrt{2} G_\mu)^{-1/2} = 246$ GeV, they can be transformed
into the Higgs basis by~\cite{LS,BS}
\be
\left(
\begin{array}{c}
H_1\\
H_2
\end{array}
\right)
=
\left(
\begin{array}{cc}
c_{\beta} & s_{\beta}\\
- s_{\beta} & c_{\beta}
\end{array}
\right)
\left(
\begin{array}{c}
\phi_1\\
\phi_2
\end{array}
\right),
\ee
where $\tan{\beta} = v_2/v_1$,
$c_\beta = \cos{\beta}$, and $s_\beta = \sin{\beta}$.
In the Higgs basis the second doublet does not get a vev
and the Goldstone bosons are in the first doublet.

Defining $\eta_3$ as the neutral imaginary component of the $H_2$ doublet, the mass eigenstates are obtained from
the three neutral states via the rotation matrix $R$
\be
\left(
\begin{array}{c}
h_1\\
h_2\\
h_3
\end{array}
\right)
= R
\left(
\begin{array}{c}
\eta_1\\
\eta_2\\
\eta_3
\end{array}
\right)
\label{h_as_eta}
\ee
which will diagonalize the mass matrix of the neutral states
via
\be
R\, {\cal M}^2\, R^T = \textrm{diag} \left(m_1^2, m_2^2, m_3^2 \right),
\ee
and $m_1 \leq m_2 \leq m_3$ are the masses of the neutral Higgs particles.
We parametrize the mixing matrix $R$ as \cite{ElKaffas:2007rq}
\be
R =
\left(
\begin{array}{ccc}
c_1 c_2 & s_1 c_2 & s_2\\
-(c_1 s_2 s_3 + s_1 c_3) & c_1 c_3 - s_1 s_2 s_3  & c_2 s_3\\
- c_1 s_2 c_3 + s_1 s_3 & -(c_1 s_3 + s_1 s_2 c_3) & c_2 c_3
\end{array}
\right)
\label{matrixR}
\ee
with $s_i = \sin{\alpha_i}$ and
$c_i = \cos{\alpha_i}$ ($i = 1, 2, 3$) and
\be
- \pi/2 < \alpha_1 \leq \pi/2,
\hspace{5ex}
- \pi/2 < \alpha_2 \leq \pi/2,
\hspace{5ex}
- \pi/2 \leq \alpha_3 \leq \pi/2.
\label{range_alpha}
\ee

The potential of the C2HDM has 9 independent parameters and we choose
as input parameters $v$, $\tan \beta$, $m_{H^\pm}$,
$\alpha_1$, $\alpha_2$, $\alpha_3$, $m_1$, $m_2$, and $\textrm{Re}(m_{12}^2)$.
The mass of the heavier neutral scalar is then given by
\be
m_3^2 = \frac{m_1^2\, R_{13} (R_{12} \tan{\beta} - R_{11})
+ m_2^2\ R_{23} (R_{22} \tan{\beta} - R_{21})}{R_{33} (R_{31} - R_{32} \tan{\beta})}.
\label{m3_derived}
\ee
The parameter space will be constrained by the condition $m_3 > m_2$.

In order to perform a study on the light Higgs bosons we need
the Higgs coupling to gauge bosons that can be written as~\cite{Barroso:2012wz}
\be
C = c_\beta R_{11} + s_\beta R_{12}
=
\cos{(\alpha_2)}\, \cos{(\alpha_1 - \beta)},
\label{C}
\ee
and the Higgs couplings to a pair of charged Higgs bosons~\cite{Barroso:2012wz}
\be
- \lambda
=
c_\beta \left[ s_\beta^2 \lambda_{145} + c_\beta^2 \lambda_3 \right] R_{11}
+
s_\beta \left[ c_\beta^2 \lambda_{245} + s_\beta^2 \lambda_3 \right] R_{12}
+
s_\beta c_\beta\, \textrm{Im}(\lambda_5)\, R_{13},
\label{lambda_hHpHm}
\ee
where $\lambda_{145} = \lambda_1 - \lambda_4 - \textrm{Re}(\lambda_5)$
and $\lambda_{245} = \lambda_2 - \lambda_4 - \textrm{Re}(\lambda_5)$.
Finally we also need the Yukawa couplings.
In order to avoid flavour changing neutral currents (FCNC)
we extend the $Z_2$ symmetry to the Yukawa Lagrangian~\cite{GWP}.
The up-type quarks couple to $\phi_2$ and the usual four models
are obtained by coupling down-type quarks and charged leptons
to $\phi_2$ (Type I) or to $\phi_1$ (Type II); or by coupling
the down-type quarks to $\phi_1$ and the charged leptons
to $\phi_2$ (Flipped) or finally by coupling the down-type
quarks to $\phi_2$ and the charged leptons to $\phi_1$ (Lepton Specific).
The Yukawa couplings can then be written, relative to the SM ones, as $a + i b \gamma_5$ with
the coefficients presented in table~\ref{tab:1}.
\begin{table}[h!]
\centering
\begin{tabular}{|lcccccccc|}
\hline
 & & Type I  & & Type II & & Lepton & & Flipped \\
 & & & & & & Specific & & \\
\hline
Up  & &
$\tfrac{R_{12}}{s_{\beta}} - i c_\beta \tfrac{R_{13}}{s_{\beta}} \gamma_5$   & &
$\tfrac{R_{12}}{s_{\beta}} - i c_\beta \tfrac{R_{13}}{s_{\beta}} \gamma_5$  & &
$\tfrac{R_{12}}{s_{\beta}} - i c_\beta \tfrac{R_{13}}{s_{\beta}} \gamma_5$   & &
$\tfrac{R_{12}}{s_{\beta}} - i c_\beta \tfrac{R_{13}}{s_{\beta}} \gamma_5$  \\*[2mm]
Down  & &
$\tfrac{R_{12}}{s_{\beta}} + i c_\beta \tfrac{R_{13}}{s_{\beta}} \gamma_5$   & &
$\tfrac{R_{11}}{c_{\beta}} - i s_\beta \tfrac{R_{13}}{c_{\beta}} \gamma_5$    & &
$\tfrac{R_{12}}{s_{\beta}} + i c_\beta \tfrac{R_{13}}{s_{\beta}} \gamma_5$   & &
$\tfrac{R_{11}}{c_{\beta}} - i s_\beta \tfrac{R_{13}}{c_{\beta}} \gamma_5$    \\*[2mm]
Leptons  & &
$\tfrac{R_{12}}{s_{\beta}} + i c_\beta \tfrac{R_{13}}{s_{\beta}} \gamma_5$   & &
$\tfrac{R_{11}}{c_{\beta}} - i s_\beta \tfrac{R_{13}}{c_{\beta}} \gamma_5$    & &
$\tfrac{R_{11}}{c_{\beta}} - i s_\beta \tfrac{R_{13}}{c_{\beta}} \gamma_5$   & &
$\tfrac{R_{12}}{s_{\beta}} + i c_\beta \tfrac{R_{13}}{s_{\beta}} \gamma_5$   \\*[2mm]
\hline
\end{tabular}
\caption{\label{tab:1} Yukawa couplings of the lightest scalar, $h_1$,
in the form $a + i b\gamma_5$.}
\end{table}

From the form of the rotation matrix $R$~\eqref{matrixR},
it is clear that when $s_2=0$, the pseudoscalar $\eta_3$
does not contribute to the mass eigenstate $h_1$. It is also
obvious that when $s_2=0$ the pseudoscalar components of all
Yukawa couplings vanish. Therefore, we can state
\ba
|s_2| = 0\ \
& \Longrightarrow &
\ \ h_1\ \textrm{is a pure scalar},
\label{pure_scalar}
\\
|s_2| = 1\ \
& \Longrightarrow &
\ \ h_1\ \textrm{is a pure pseudoscalar}.
\label{pure_pseudoscalar}
\ea

There are however other interesting scenarios that could be in principle allowed.
We could ask ourselves if a situation where the scalar couplings $a_{F} \approx 0$ ($F=U,D,L$)
is still allowed after the 8 TeV run. As $a_U$ is fixed (the same for all Yukawa types) and given by
$a_U = R_{12}/s_{\beta} = s_1 c_2/s_{\beta}$, it can only be small if $s_1 \approx 0$.
If instead $c_2 \approx 0$ the $h_1 V V$ coupling $C$ in eq.~\eqref{C} would
vanish which is already disallowed by experiment. There is one other coupling
that could also vanish, which is $R_{11}/c_{\beta} = c_1 c_2/c_{\beta}$ (this is
for example the expression for $a_D$ in Type II). Again this scalar part could
vanish if $c_1 \approx 0$. In either case, $s_1 \approx 0$ or $c_1 \approx 0$,
the important point to note is that the pseudoscalar component of the 125 GeV Higgs is not constrained
by the choice of $\alpha_1$ because it depends only on $s_2$. We will discuss these scenarios in detail
in section~\ref{zero}.

\section{Present status of the C2HDM}
\label{status}

We start by briefly reviewing the status of the C2HDM after the 8 TeV run. We will
generate points in parameter space with the following conditions:
the lightest neutral scalar is $m_1 = 125$ GeV~\footnote{The
latest results on the measurement of the Higgs mass are $125.36 \pm 0.37$ GeV
from ATLAS~\cite{Aad:2014aba} and $125.02 +0.26 -0.27$ (stat) $+0.14 -0.15$ (syst) GeV
from CMS~\cite{Khachatryan:2014jba}.},
the angles $\alpha_{1,2,3}$ all vary in the interval $[-\pi/2, \, \pi/2]$,
$1 \leq \tan{\beta} \leq 30$, $ m_1 \leq m_2 \leq 900\, \textrm{GeV}$ and
$-(900\, \textrm{GeV})^2 \leq Re(m_{12}^2) \leq (900\, \textrm{GeV})^2$.
Finally, we consider different ranges for the charged Higgs mass because
the constraints from B-physics, and in particular
the ones from $b \ra s \gamma$, affect differently Type II/F and Type I/LS.
In Type II and F the range for the charged Higgs mass is
$340\, \textrm{GeV} \leq m_{H^\pm} \leq 900\, \textrm{GeV}$ due to
$b \ra s \gamma$ which forces $\mhpm \gtrsim 340$ GeV almost independently of $\tan \beta$~\cite{BB}.
In Type I and LS the range is $100\, \textrm{GeV} \leq m_{H^\pm} \leq 900\, \textrm{GeV}$ because
the constraint from $b \ra s \gamma$ is not as strong. The remaining constraints from
B-physics~\cite{Deschamps:2009rh, gfitter1} and from the
$R_b\equiv\Gamma(Z\to b\bar{b})/\Gamma(Z\to{\rm hadrons})$~\cite{Ztobb} measurement
have a similar effect on all models forcing $\tan{\beta} \gtrsim 1$.
The choice of the lower bound of 100 GeV is due to LEP searches on $e^+ e^- \to H^+ H^-$~\cite{Abbiendi:2013hk}
and the latest LHC results on $pp \to \bar t \, t (\to H^+ \bar b )$~\cite{ATLASICHEP, CMSICHEP}.
Very light neutral scalars are also constrained by LEP results~\cite{lepewwg}.

All points comply to the following theoretical constraints: the potential is bounded from
below~\cite{Deshpande:1977rw}, perturbative unitarity is
enforced~\cite{Kanemura:1993hm, Akeroyd:2000wc,Ginzburg:2003fe} and all allowed points conform
to the oblique radiative parameters~\cite{Peskin:1991sw, Grimus:2008nb, Baak:2012kk}.

The signal strength is defined as
\begin{equation}
\mu^{h_i}_f \, = \, \frac{\sigma \, {\rm BR} (h_i \to
  f)}{\sigma^{\scriptscriptstyle {\rm SM}} \, {\rm BR^{\scriptscriptstyle{\rm SM}}} (h_i \to f)}
\label{eg-rg}
\end{equation}
where $\sigma$ is the Higgs boson production cross section and ${\rm BR} (h_i \to f)$ is
the branching ratio of the $h_i$ decay into the final state $f$;  $\sigma^{\scriptscriptstyle {\rm {SM}}}$
and ${\rm BR^{\scriptscriptstyle {\rm SM}}}(h \to f)$ are the respective quantities calculated in the SM.
The gluon fusion cross section is calculated at NNLO using HIGLU~\cite{Spira:1995mt} together with the
corresponding expressions for the CP-violating model in~\cite{Fontes:2014xva}.
SusHi \cite{Harlander:2012pb} at NNLO is used for calculating $b \bar{b} \ra h$, while
$Vh$ (associated production), $t \bar{t} h$ and $VV \ra h$ (vector boson fusion) can be found
in~\cite{LHCCrossSections}. As previously discussed we will consider the rates
for the processes $\mu_{VV}$, $\mu_{\gamma \gamma}$ and $\mu_{\tau \tau}$ to be within $20$\%
of the expected SM value, which at present roughly matches the average precision at $1\sigma$. It was shown in
\cite{Fontes:2014xva}
that taking into account other processes with the present attained precision has no
significant impact in the results.

\begin{figure}[h!]
\centering
\includegraphics[width=0.46\linewidth]{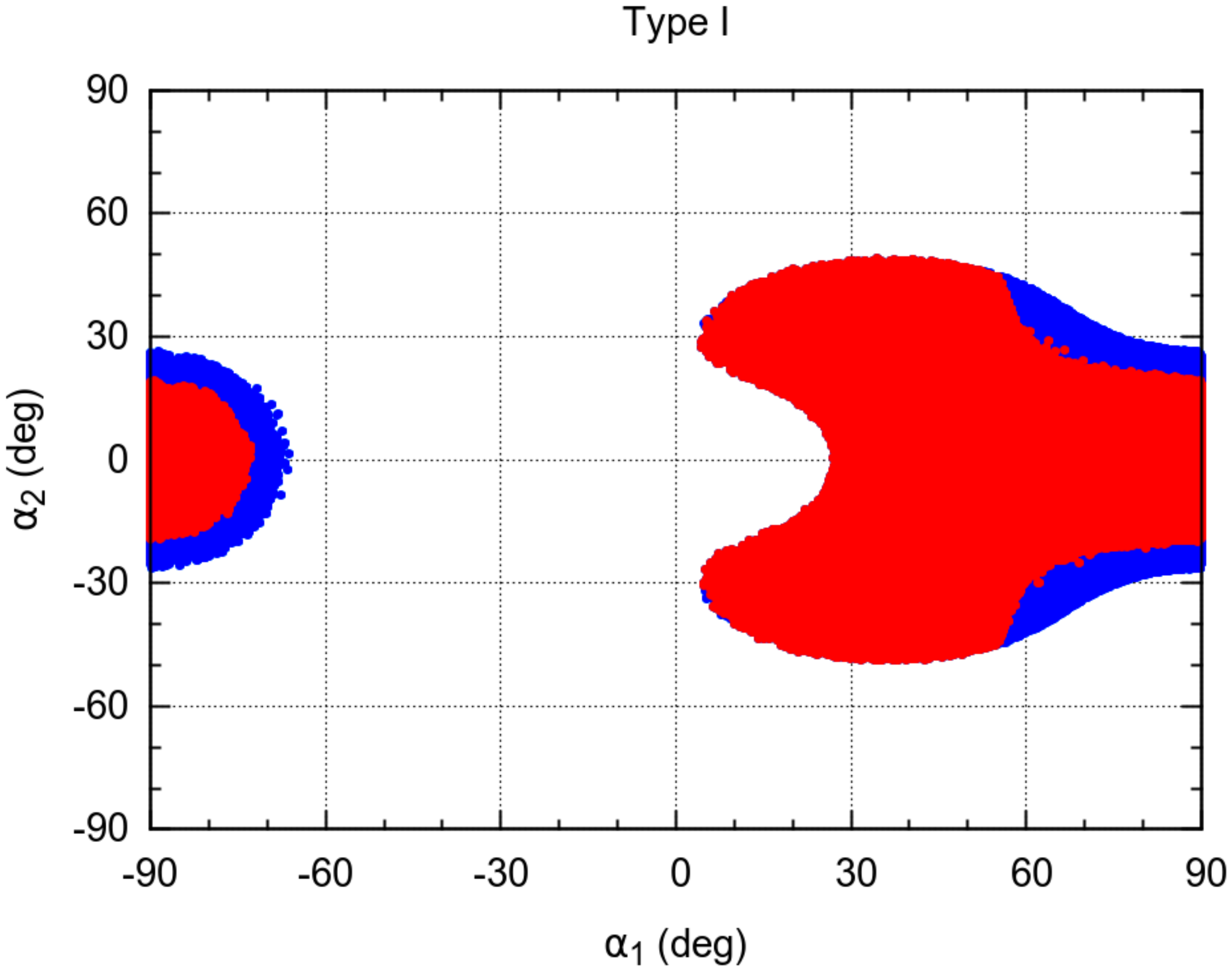}
\hspace{0.02\linewidth}
\includegraphics[width=0.46\linewidth]{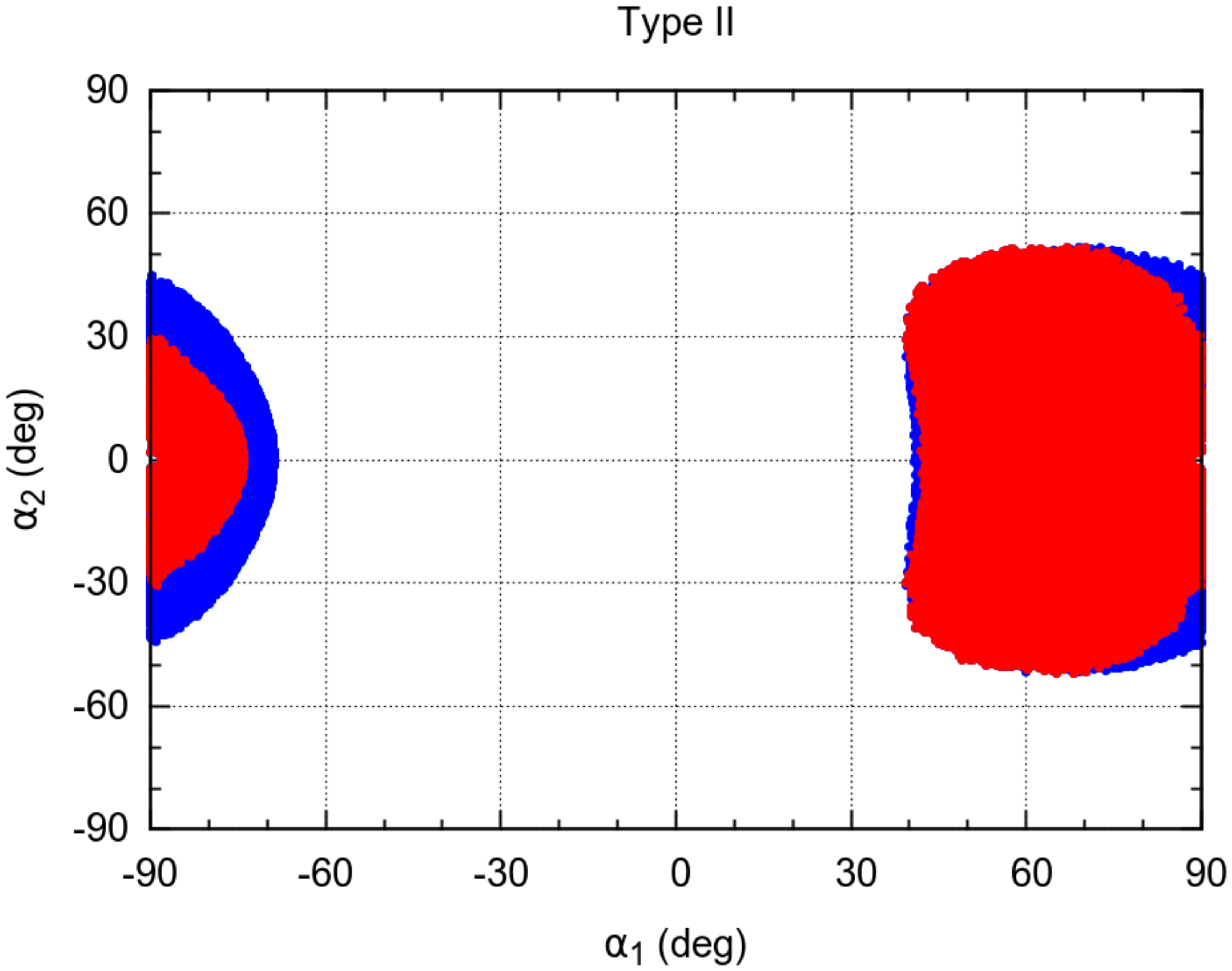}
\caption{$\alpha_2$ vs. $\alpha_1$ for Type I (left) and Type II (right).
The rates are taken to be within 20$\%$ of the SM predictions. The colours are superimposed with
cyan/light-grey for $\mu_{VV}$, blue/black for $\mu_{\tau \tau}$ and finally red/dark-grey for $\mu_{\gamma \gamma}$ with
a center of mass energy of 8 TeV.}
\label{fig:F1}
\end{figure}
%
We start by examining the parameter space for a center of mass energy of $\sqrt{s} = 8$ TeV
corresponding to the end of the first LHC run. The rates are taken at $20$\% and the effect
of considering each of the rates at a time is shown by superimposing the colours,
cyan/light-grey ($\mu_{VV}$),
blue/black ($\mu_{\tau \tau}$) and finally red/dark-grey ($\mu_{\gamma \gamma}$).
In figure~\ref{fig:F1} we present the allowed space for the angles $\alpha_2$
vs. $\alpha_1$ for Type I (left) and Type II (right)
with all theoretical and
collider constraints taken into account. The corresponding plots for the Flipped (Lepton specific) are
very similar to the one for Type II (I) and are not shown.
It was expected
that $\alpha_2$ would be centred around zero where the pseudoscalar component vanishes. Also $\alpha_1$ plays
the role of $\alpha + \pi/2$, where $\alpha$ is the rotation angle in the CP-conserving case
\footnote{We can choose a parametrization where the angle $\alpha_1$ is exactly $\alpha$ in that limit. See for example the
definition of the rotation matrix in~\cite{Inoue:2014nva} as compared to our equation~\ref{matrixR}.}.
In previous works for the CP-conserving model~\cite{Ferreira:2012nv, Fontes:2014tga}
we have made estimates for the allowed parameters
based on the assumption that the production is dominated by gluon fusion and that
$\Gamma(h_1 \to b \bar b)$ is to a good approximation the Higgs total width.
Under a similar approximation, we can write
for Type I and large $\tan \beta$ (when $\tan \beta \gg 1$, $b_i \ll 1$ and we recover the CP-conserving
Yukawa couplings)
\begin{equation}
\mu_{VV}^{I} \approx \cos^2 \alpha_2 \, \cos^2 (\beta - \alpha_1)  \, .
\label{eq:muI}
\end{equation}
Since we are considering a $20$\% accuracy, it is clear that neither $\cos \alpha_2$
nor $\cos (\beta - \alpha_1)$ can be close to zero. In fact, a measurement of $\mu_{VV}$
with a $20$\% ($5$\%) accuracy and centred at the SM expected value implies $\cos^2 \alpha_2 \gtrsim 0.8 (0.95)$
and consequently $|\sin \alpha_2| \lesssim 0.45 (0.22)$ and $|\alpha_2| \lesssim 27 \degree (13 \degree)$.
Although the approximations captures the features, the plot does not reproduce the exact value
of the limit, which for a $20$\% accuracy is slightly below $50 \degree$.


\begin{figure}[h!]
\centering
\includegraphics[width=0.33\linewidth]{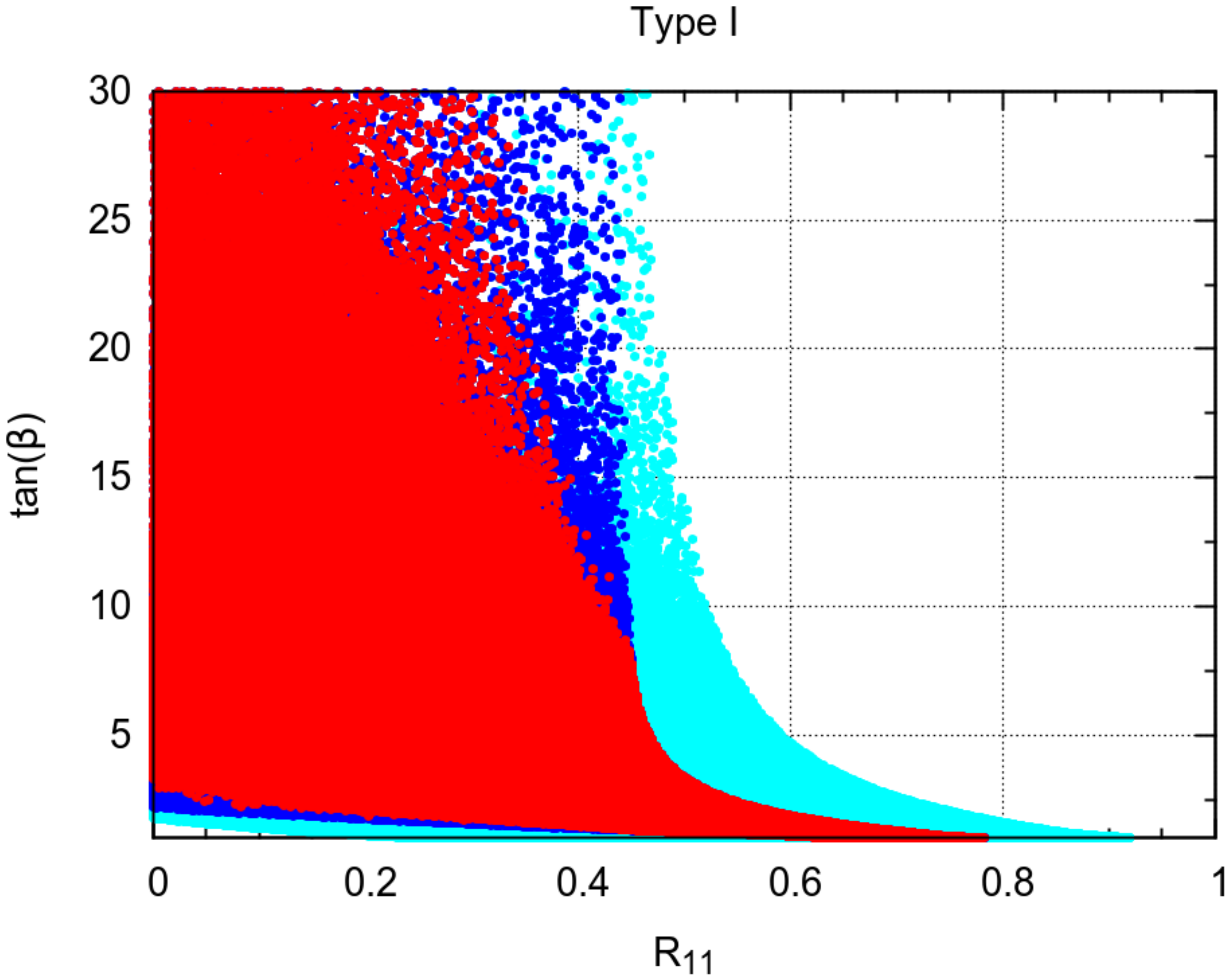}
\hspace{-0.02\linewidth}
\includegraphics[width=0.33\linewidth]{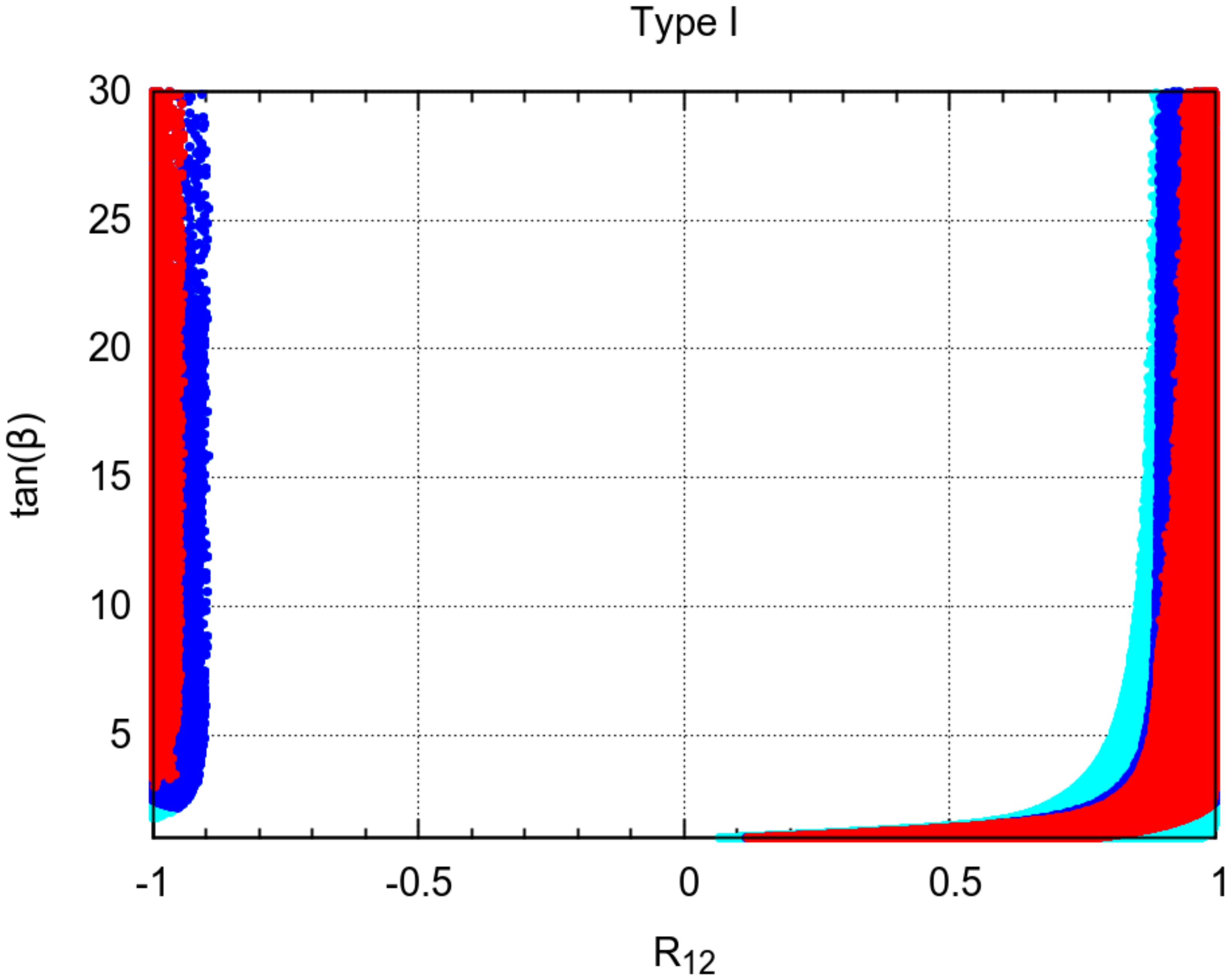}
\hspace{-0.02\linewidth}
\includegraphics[width=0.33\linewidth]{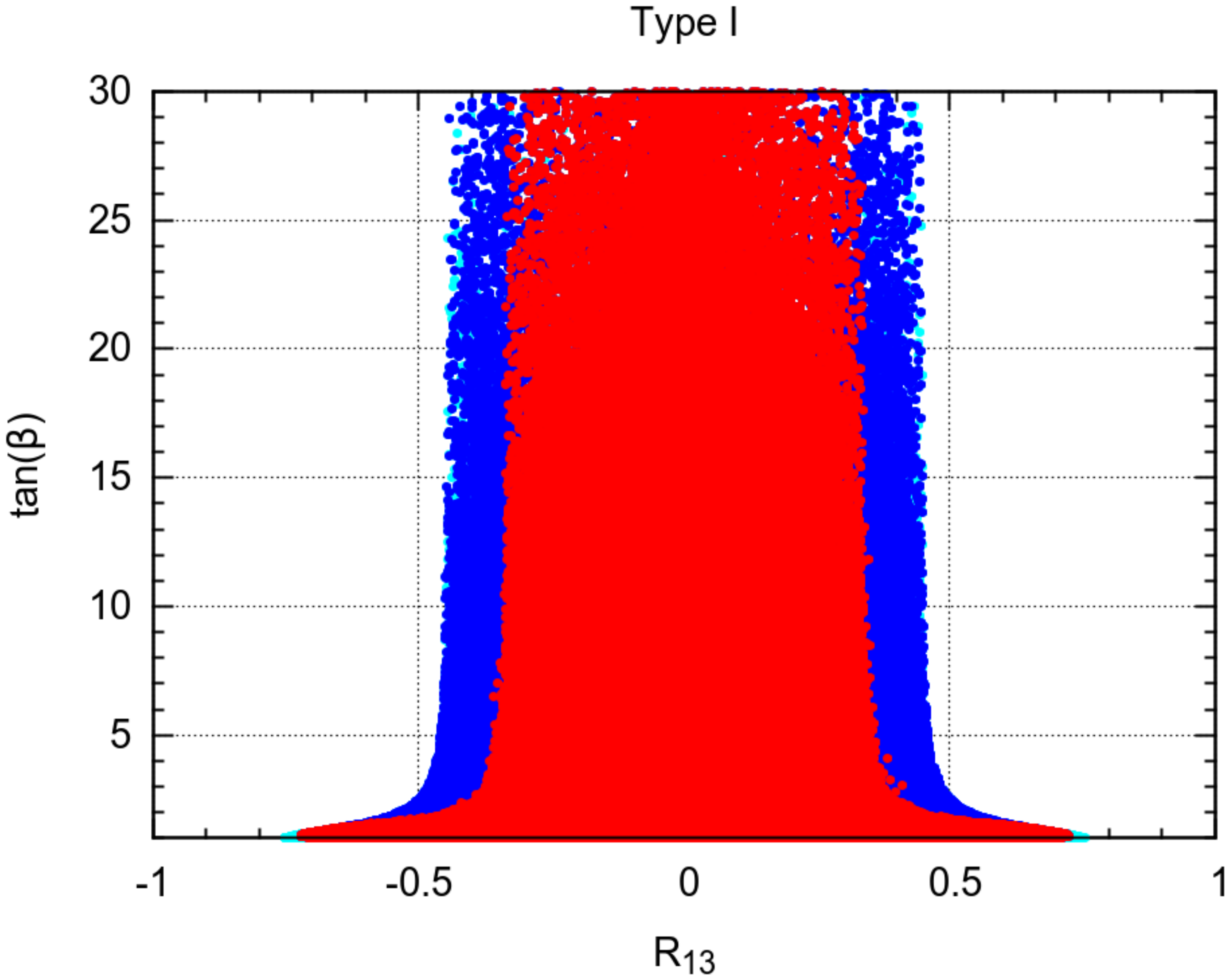}
\vspace{1mm}\\
\includegraphics[width=0.33\linewidth]{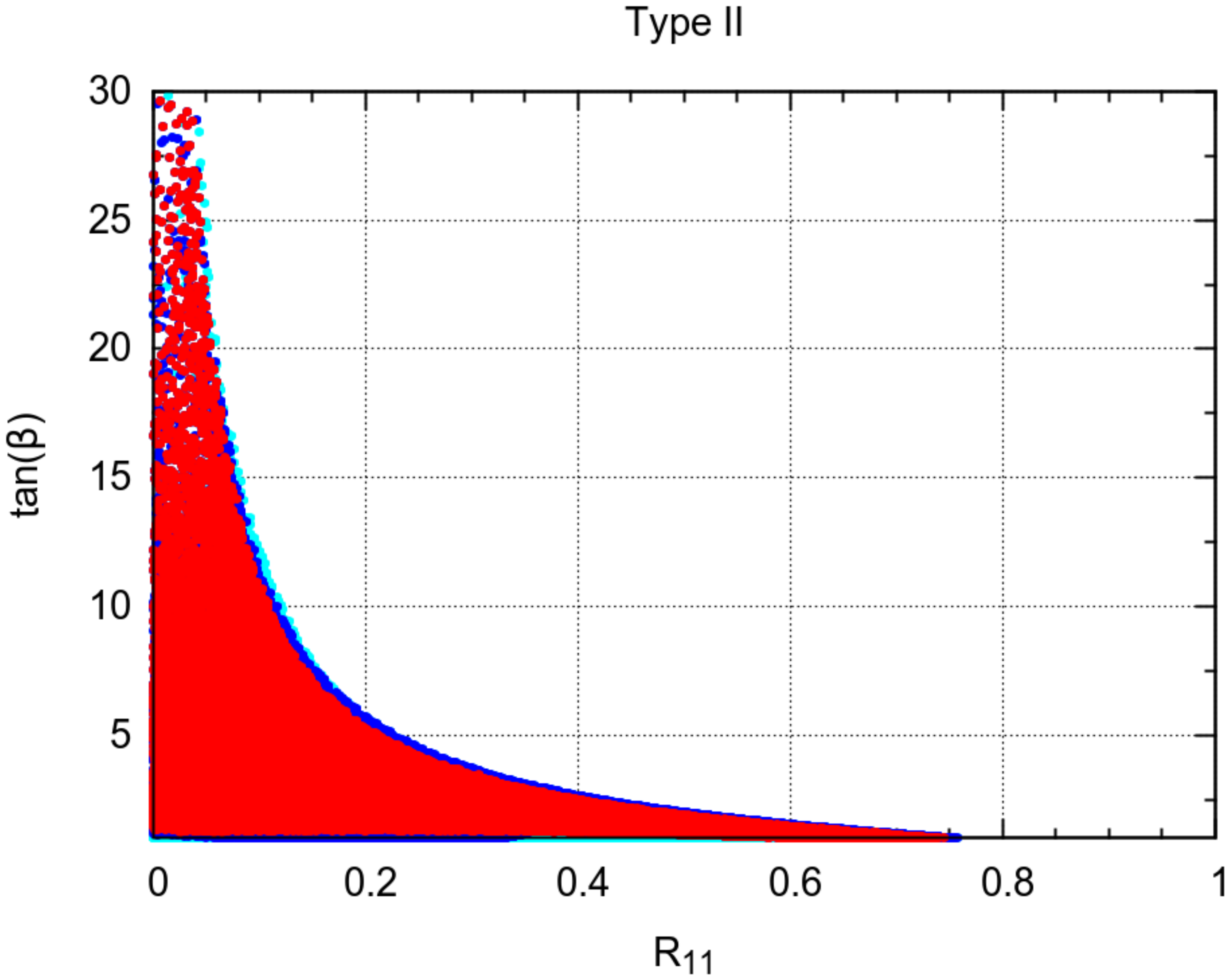}
\hspace{-0.02\linewidth}
\includegraphics[width=0.33\linewidth]{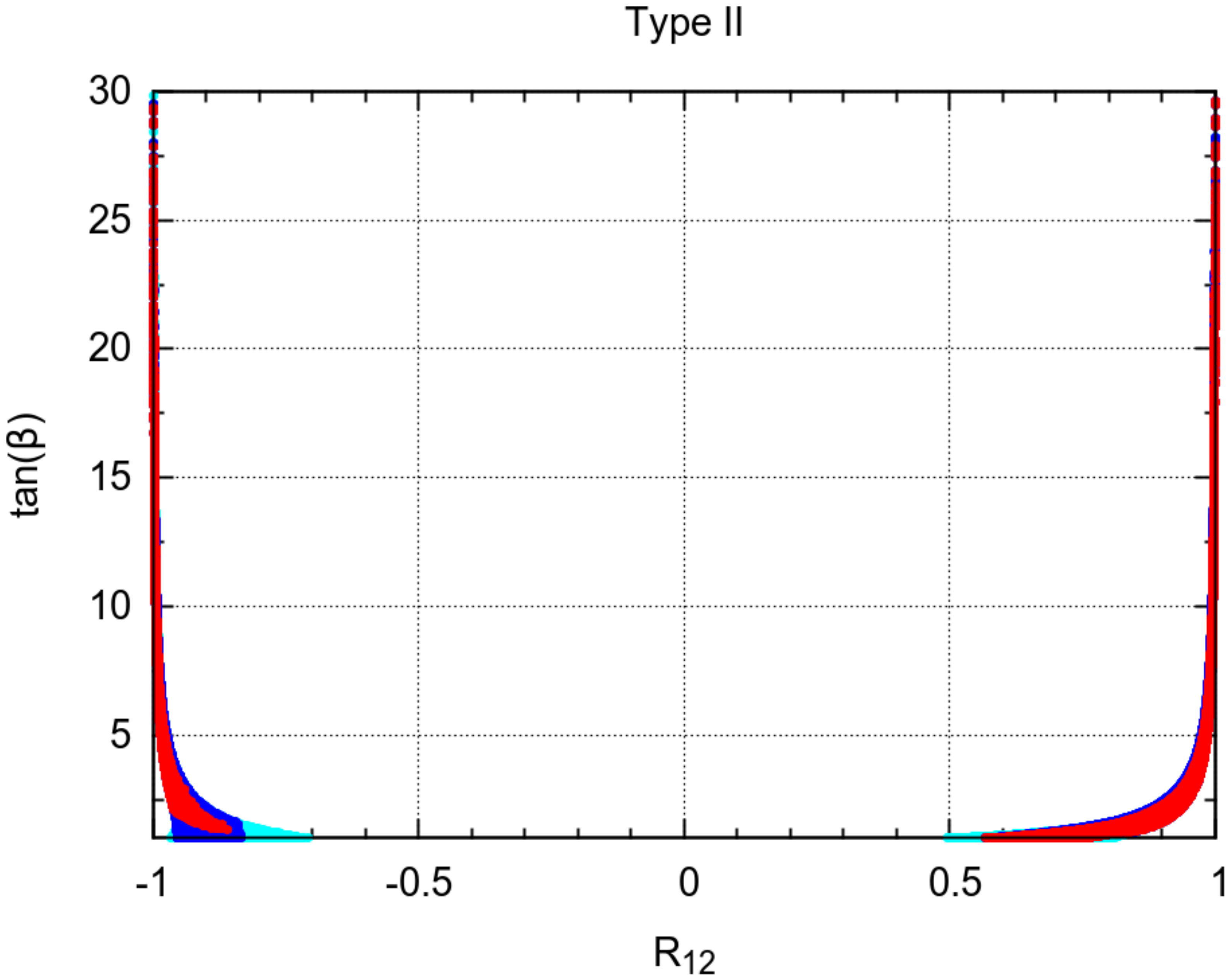}
\hspace{-0.02\linewidth}
\includegraphics[width=0.33\linewidth]{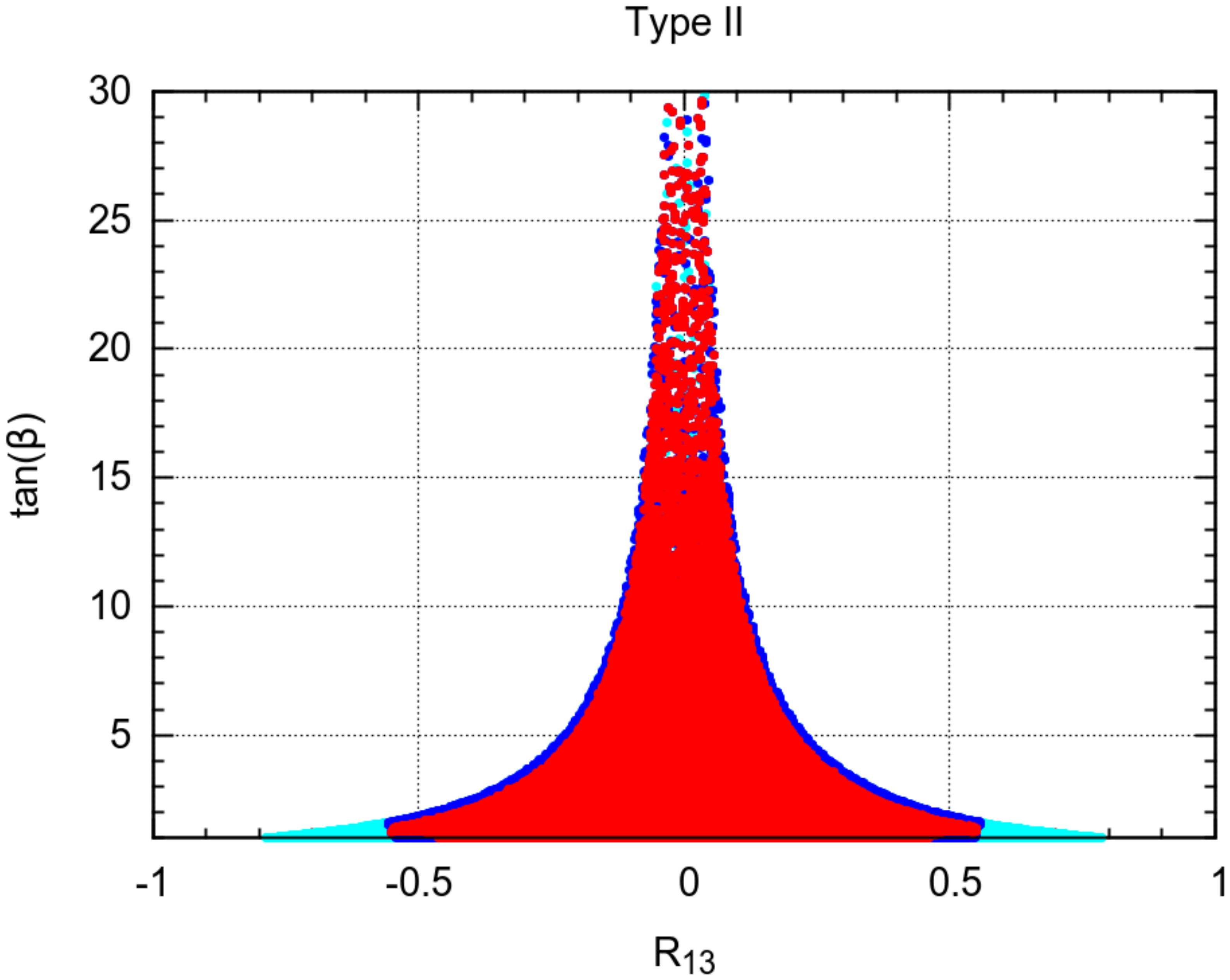}
\caption{Top: $\tan \beta$ as a function of $R_{11}$ (left), $R_{12}$ (middle)
and $R_{13}$ for Type I. Bottom: same but for Type II. The rates are taken to be
within 20$\%$ of the SM predictions. The colours are superimposed with
cyan/light-grey for $\mu_{VV}$, blue/black for $\mu_{\tau \tau}$ and finally red/dark-grey for $\mu_{\gamma \gamma}$ with
a center of mass energy of 8 TeV.}
\label{fig:F2}
\end{figure}
%
In figure~\ref{fig:F2} we show $\tan \beta$ as a function of $R_{11}$ (left), $R_{12}$ (middle)
and $R_{13}$ (right). The upper plots are for Type I and the lower plots for Type II. Again the differences
of Type II (I) relative to F (LS) are small and we do not show the corresponding plots.
We start with $R_{13}$ which is just $\sin \alpha_2$, thus measuring the amount of CP-violation
for the 125 GeV Higgs, that is, the magnitude of its pseudoscalar component.
The allowed points are centred around zero where we recover a SM-like Higgs
Yukawa coupling for the lightest scalar state.
The differences between the models only occur for large $\tan \beta$, reflecting the different
angle dependence of the couplings in the various models. We now discuss
$R_{12}=\sin \alpha_1 \cos \alpha_2$. Using the same approximation for $\mu_{VV}$ as in eq.(\ref{eq:muI})
we can write for large $\tan \beta$
\begin{equation}
\mu_{VV}^{I} \approx R_{12}^2  \, ,
\label{eq:muI2}
\end{equation}
which means that if we take $|R_{12}^2| > 0.8$ then $R_{12}>0.89$ or $R_{12}< - 0.89$. These are exactly
the bounds we see in the plots for Type I. Therefore, as already happened for the CP-conserving case
it is mainly $\mu_{VV}$ that constrains $|R_{12}|$ to be close to $1$ especially for large $\tan \beta$.
Finally $R_{11}=\cos \alpha_1 \cos \alpha_2$ is only indirectly constrained by the bounds on $\alpha_1$
and $\alpha_2$. Since the pure scalar part of the coupling relative to the SM
is proportional to $R_{11}^2 \, (1+\tan^2 \beta)$ it is natural that when $R_{11}$ increases,
$\tan \beta$ decreases. However, the most important point to note is that $R_{11} = 0$ is allowed.
Although $R_{11}$ is never part of the Yukawa couplings in Type I, it appears in pure scalar couplings for
down-type quarks or/and charged leptons in the remaining types. This in turn implies that scenarios
where $a_D =0$ and/or $a_L =0$ are not excluded. Models Type II, F and LS can therefore have a pure
pseudoscalar component for some of its Yukawa couplings. This scenario will be discussed in detail
in the next section.

\begin{figure}[h!]
\centering
\includegraphics[width=0.33\linewidth]{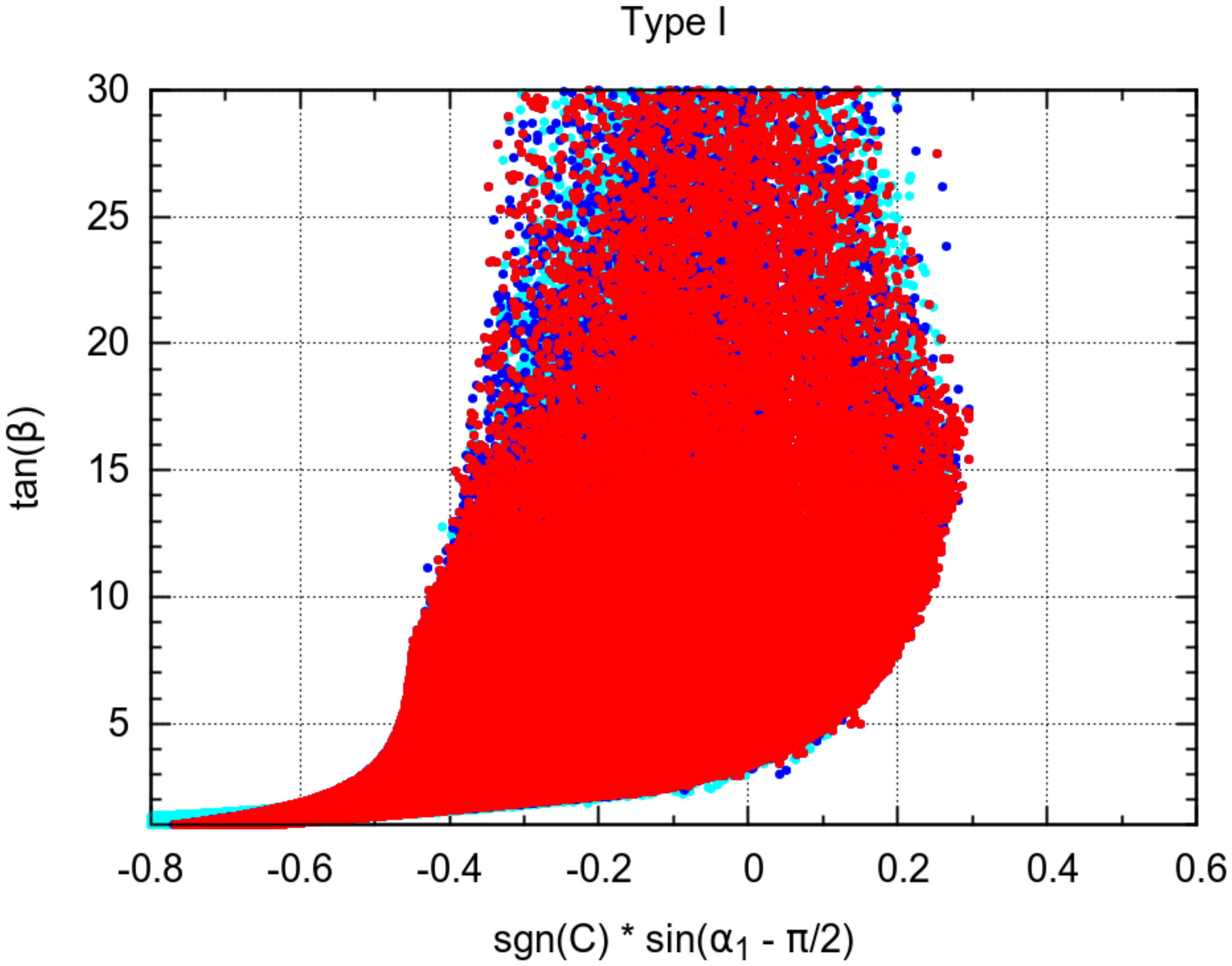}
\hspace{-0.02\linewidth}
\includegraphics[width=0.33\linewidth]{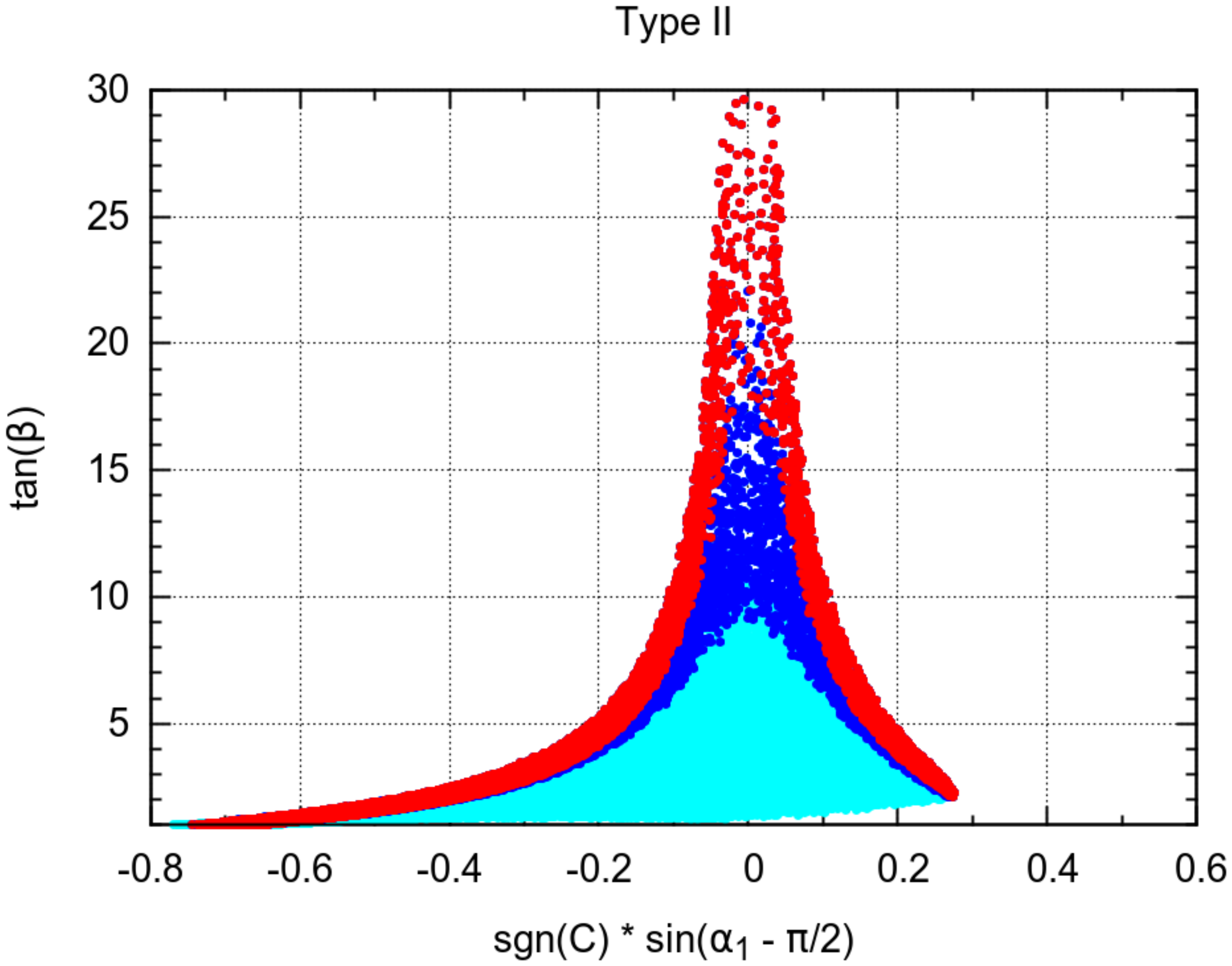}
\hspace{-0.02\linewidth}
\includegraphics[width=0.33\linewidth]{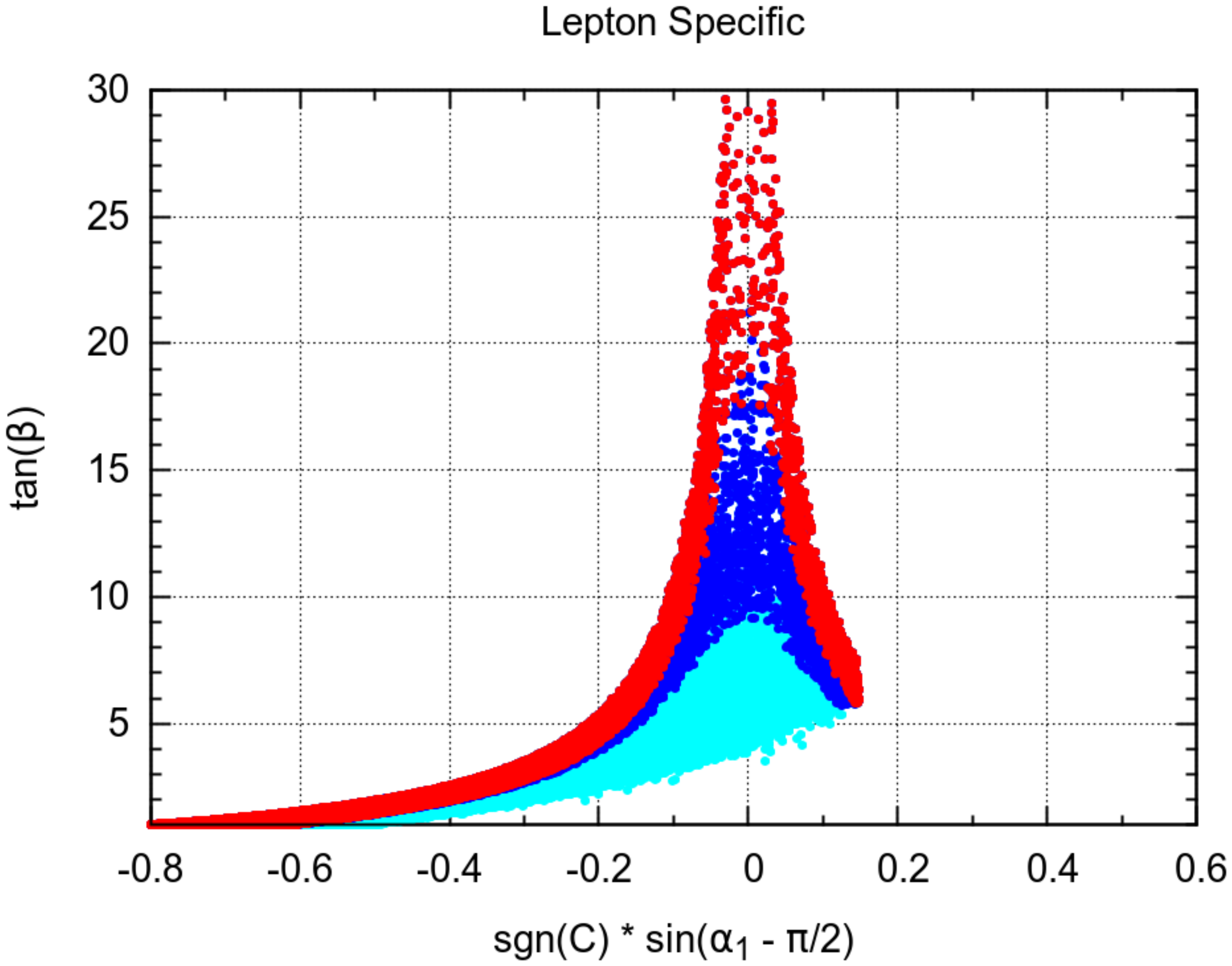}
\caption{$\tan \beta$ as a function of $\sin (\alpha_1-\pi/2)$
with all rates at $20$\% for Type I (left), Type II (middle) and LS (right).
All angles are free to vary in their allowed range (cyan/light-grey) and we impose
the constraint $s_2 < 0.1$ (blue/black) and $s_2 < 0.05$ (red/dark-grey).}
\label{fig:F3}
\end{figure}
%
In figure~\ref{fig:F3} we present $\tan \beta$ as a function of $\sin (\alpha_1-\pi/2)$
with all rates at $20$\% for Type I (left), Type II (middle) and LS (right).
All angles are free to vary in their allowed range and we present scenarios
for which $s_2 < 0.1$ and $s_2 < 0.05$. We plot  $\alpha_1-\pi/2$ instead of $\alpha_1$
to match the usual definition for the CP-conserving model. Since
we recover the CP-conserving $h_1$ couplings when
$s_2 =0$, the red/dark-grey outer layer for Type II and LS
has to match the bounds for the angle $\alpha$ in the CP conserving case which is
indeed the case~\cite{ourreview}. If we identify $\alpha_1$ with $\alpha+\pi/2$, where $\alpha$
is the rotation angle for the CP-conserving scenario, we can write the coupling
to gauge bosons as
\be
g_{hVV}^{\scriptscriptstyle {\rm CPV}} = \cos{(\alpha_2)} \, g_{hVV}^{\scriptscriptstyle {\rm CPC}} \, .
\label{C1}
\ee
Hence, for Type I $\mu_{VV}$ will either give the same bound as in the CP-conserving case or
worse as $\cos{(\alpha_2)}$ decreases. However, for Type II, the same approximation
that lead to eq. (\ref{eq:muI}) for Type I results for Type II in
\begin{equation}
\mu_{VV}^{II} \approx \frac{ \cos^2 \alpha_2 \, \cos^2 (\beta - \alpha_1)}{\tan^2 \beta} \, \, \,
\frac{\sin^2 \alpha_1 \, \cos^2 \alpha_2 + \sin^2 \alpha_2 \, \cos^2 \beta}
{\cos^2 \alpha_1 \, \cos^2 \alpha_2 + \sin^2 \alpha_2 \, \sin^2 \beta} \, .
\label{approxtype2}
\end{equation}
Again, if $s_2=0$ we recover the CP-conserving expression. However, it can be shown that
larger values of $s_2$ together with smaller values of $\tan \beta$ still fulfil the constraints
on the rates. We conclude that in Type I, the allowed parameter space is the same as in the
CP-conserving case while, for the remaining types and for a given $\alpha_1$, the upper bound on $\tan \beta$
is the same as in the CP-conserving case. But, now, there is no lower bound on $\tan \beta$.

\section{The zero scalar components scenarios and the LHC run 2}
\label{zero}

In the previous section we have shown that $R_{11}=0$ is still allowed, which implies that
the pure scalar components of the Yukawa couplings can be zero in some scenarios. This
possibility arises in Type II, F and LS. In particular for Type II we can have either
$a_D = 0$ or $a_L = 0$ while in F (LS) only $a_D = 0$ ($a_L = 0$) is possible.
For definiteness let us now analyse the case where $a_D = 0$ in Type II.
Since $a_D = R_{11}/c_{\beta} = c_1 c_2/c_{\beta}$
we could in principle have $c_1=0$ or $c_2=0$. However, $c_2=0$ would mean that the gauge
bosons would not couple at tree level to the Higgs, a scenario that is
ruled out by experiment as shown in the previous section.
Setting $c_1=0$ we get, in Type II, $a_D=a_L =0$ and
\begin{equation}
a_U^2=c_2^2/s_\beta^2, \quad b_U^2=s_2^2/t_\beta^2, \quad b_D^2=b_L^2=t_{\beta}^2 s_2^2, \quad C^2=s_\beta^2 c_2^2 \,.
\end{equation}
%

%
\begin{figure}[h!]
\centering
\includegraphics[width=0.49\linewidth]{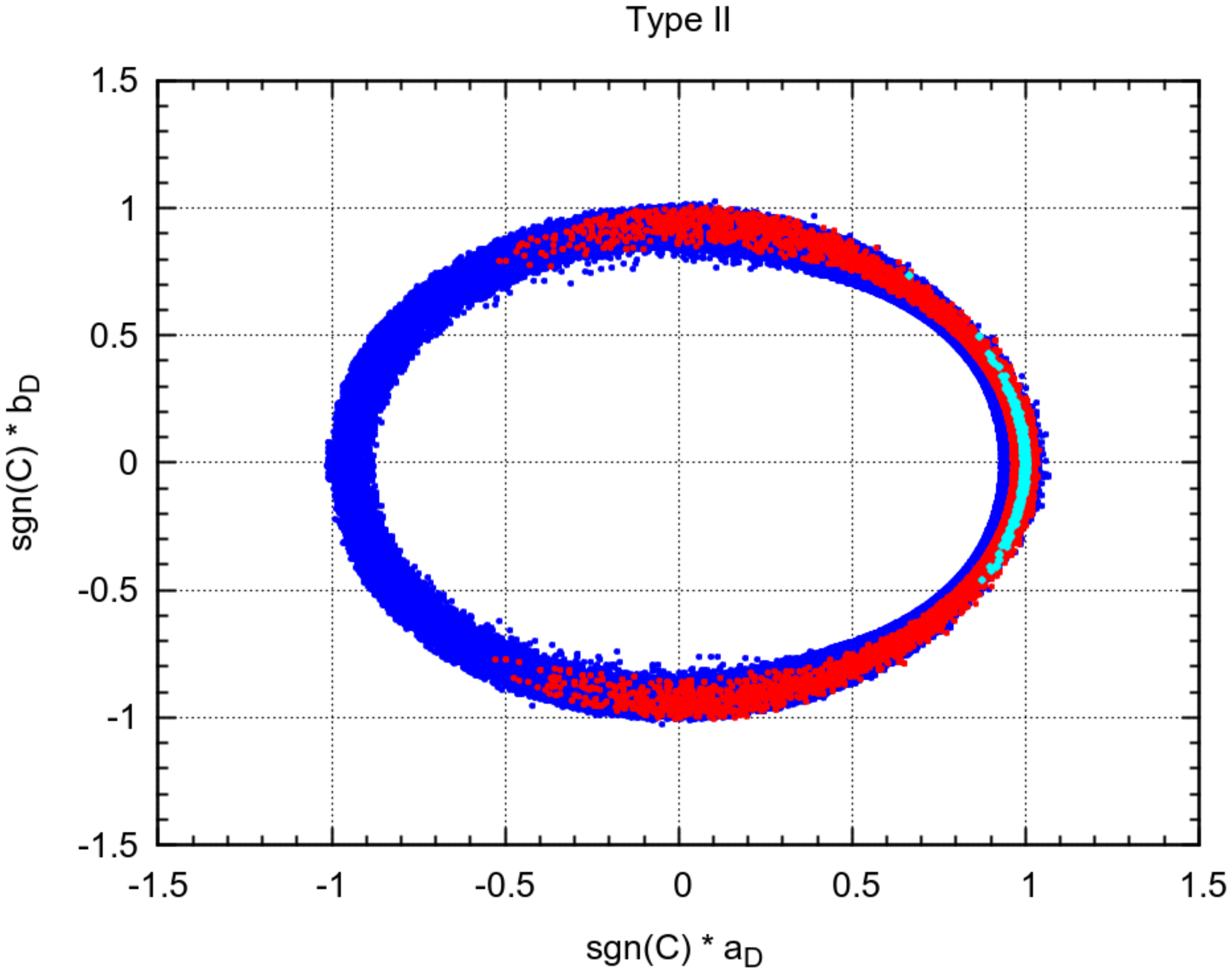}
\hspace{-0.02\linewidth}
\includegraphics[width=0.49\linewidth]{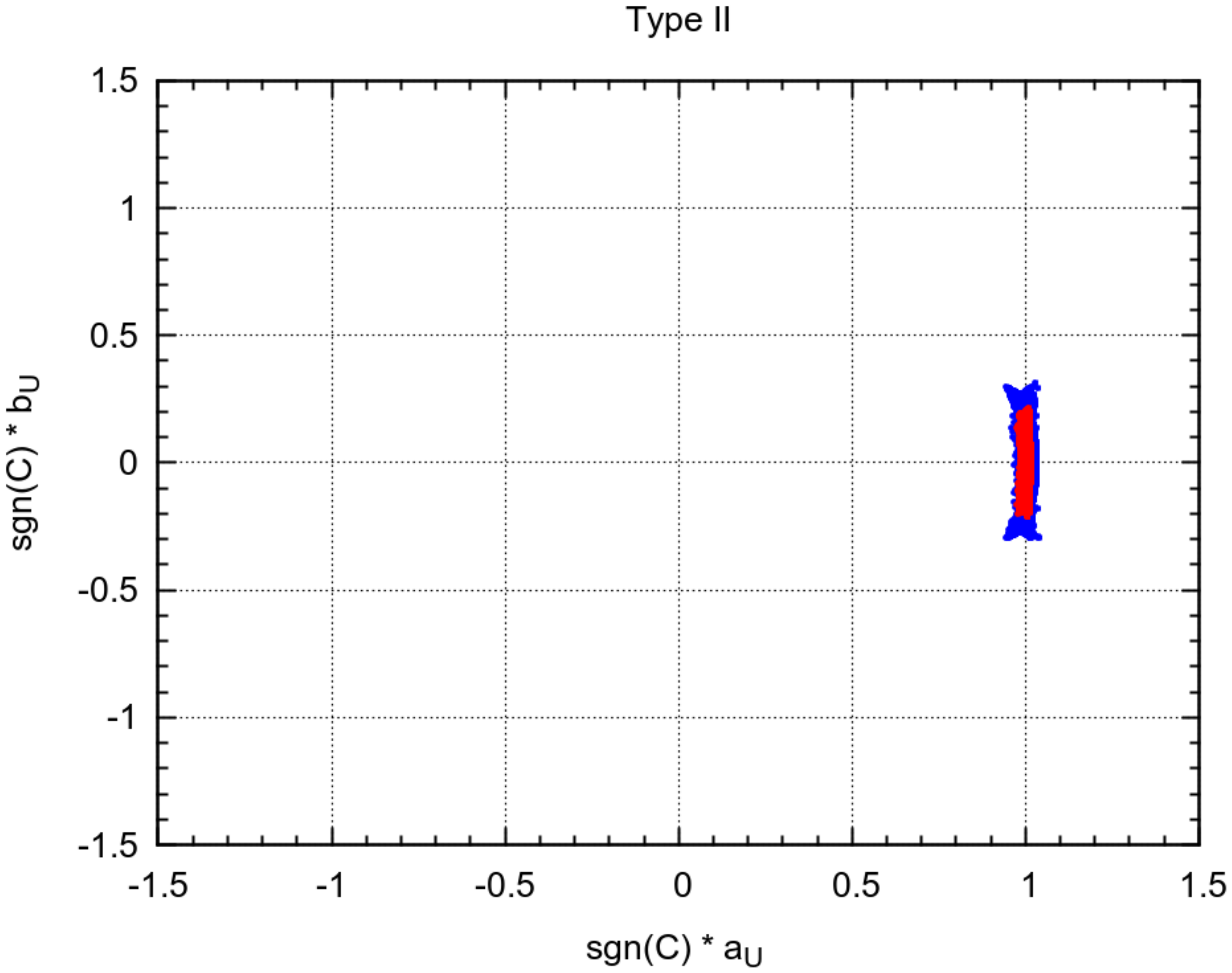}
\caption{Left: sgn$(C)$ $b_D$ $=$ sgn$(C)$ $b_L$ as a function of sgn$(C)$ $a_D$ $=$ sgn$(C)$ $a_L$  for Type II and a center of mass
energy of $13$ TeV with all rates
at $10$\% (blue/black), $5$\% (red/dark-grey), and $1$\% (cyan/light-grey).
Right: same, but for sgn$(C)$ $b_U$ as a function of sgn$(C)$ $a_U$.}
\label{fig:F4}
\end{figure}
%
In the left panel of figure~\ref{fig:F4} we show $b_D=b_L$ as a function
of $a_D=a_L$ for Type II and a center of mass energy of $13$ TeV with all rates
at $10$\% (blue/black), $5$\% (red/dark-grey), and $1$\% (cyan/light-grey) (in order to avoid the dependence
on the phase conventions in choosing the range for the angles $\alpha_i$,
we plot sgn$(C) \,a_i$ (sgn$(C) \,b_i$) instead of $a_i$ ($b_i$) with $i=U,D,L$).
It is quite interesting
to note that this scenario is still possible with the rates at $5$\% of the SM value
at the LHC at 13 TeV. We have
checked that this is still true at $2$\% and only when the accuracy reaches
$1$\% are we able to exclude the scenario. So far we have discussed $a_D=0$.
Another interesting point is that
when  $|a_D| \to 0$, $|b_D| \to 1$. The requirement that $|b_D| \approx 1$
implies that the couplings of the up-type quarks to the lightest Higgs take the form
\begin{equation}
a_U^2=( 1 -s_2^4)=(1-1/t_\beta^4), \quad b_U^2=s_2^4=1/t^4_\beta,
\end{equation}
while the coupling to massive gauge bosons is now
\begin{equation}
C^2 = (t^2_\beta-1)/(t^2_\beta+1) = (1 - s_2^2)/(1 + s_2^2)  \, .
\end{equation}
In the right panel of figure~\ref{fig:F4} we now show $b_U$ as a function
of $a_U$ for Type II with the same colour code.
We conclude from the plot that the constraint on the values of $(a_U,\, b_U)$ are already quite
strong and will be much stronger in the future just taking into account the
measurement of the rates.

\begin{figure}[h!]
\centering
\includegraphics[width=0.49\linewidth]{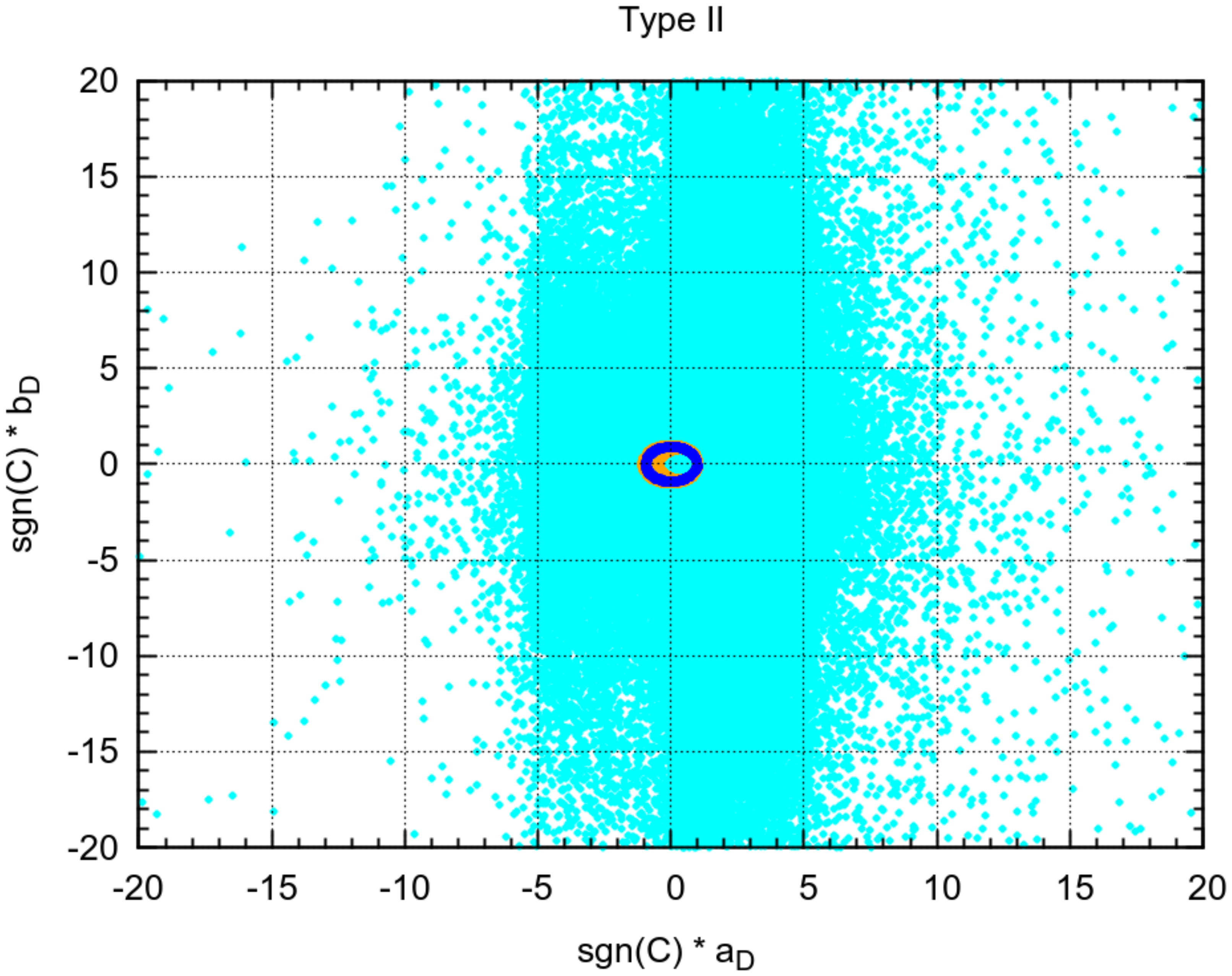}
\hspace{-0.02\linewidth}
\includegraphics[width=0.49\linewidth]{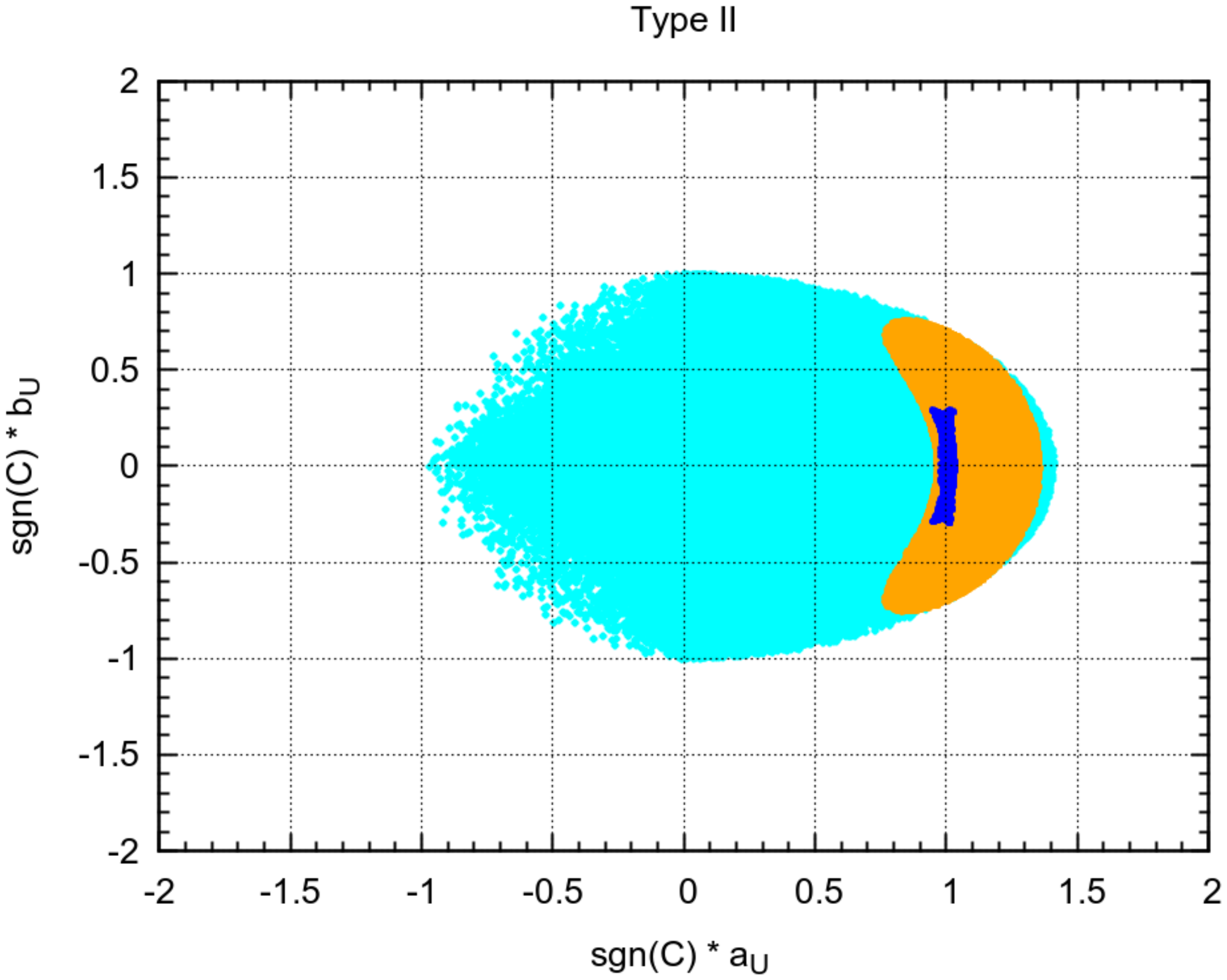}
\caption{Left: Simulation points on the sgn$(C)\, a_D$ versus sgn$(C)\, b_D$
plane. In cyan/light-grey (orange/dark-grey; blue/black) we show
the points which pass all theoretical constraints
(pass, in addition, the restriction from $\mu_{VV}$ at 10\%;
pass, in addition, the restriction from all $\mu$'s at 10\%).
Right: Same constraints on the sgn$(C)\, a_U$ versus sgn$(C)\, b_U$
plane.}
\label{fig:NEW}
\end{figure}

We would like to understand why $a_U \sim 1$ and $b_U \sim 0$,
while the bounds are much looser for $a_D$.
We start by noting that the couplings impose different constraints
on the up and down sectors.
Indeed,
from table~\ref{tab:1} and eq.~\eqref{C},
\begin{equation}
R_{11} =\frac{C- s_\beta^2 \, a_U}{c_\beta}; \quad R_{12} = s_\beta \, a_U; \quad R_{13} = - \tan{\beta} \, b_U,
\label{RsUP}
\end{equation}
for the up sector,
while
\begin{equation}
R_{11} = c_\beta \, a_D; \quad R_{12} = \frac{C- c_\beta^2 \, a_D}{s_\beta}; \quad R_{13} = -  \frac{c_\beta}{s_\beta}\, b_D,
\label{RsDOWN}
\end{equation}
for the down sector.
In the first case,
$R_{12}^2 + R_{13}^2 < 1$ leads
to
\be
a_U^2 + \frac{b_U^2}{c_\beta^2} < \frac{1}{s_\beta^2}.
\label{eq:elliUP1}
\ee
Noting that the $\tan{\beta} > 1$ constraint forces
$c_\beta < 1/\sqrt{2}$ and $s_\beta > 1/\sqrt{2}$,
we find
$b_U < 1$, while $a_U < \sqrt{2}$.
This is what we see in the right panel of figure~\ref{fig:NEW},
where in cyan we show points which are subject
only to the theoretical constraints.
%
%
We see that all points lye inside the ellipse in eq.~\eqref{eq:elliUP1}.
The constraint from the $\mu_{VV}$ bound
(orange/dark-grey points in the right panel of
figure~\ref{fig:NEW}) then places the points
on a section of that ellipse close to $(a_U, b_U) \sim (0,1)$.

The situation is completely different for the down sector.
Indeed,
a similar analysis starting from eqs.~\eqref{RsDOWN}
and $R_{11}^2 + R_{13}^2 < 1$, would lead to
\be
a_D^2 + \frac{b_D^2}{s_\beta^2} < \frac{1}{c_\beta^2}.
\label{eq:elliDOWN1}
\ee
Since $c_\beta$ can be very small, this entails no constraint at all,
agreeing with the fact that the cyan/light-grey points
in the left panel of figure~\ref{fig:NEW} have no restriction.
In contrast,
it is the bound on $\mu_{VV}$ which constrains the parameter
space to the orange/dark-grey circle centered at $(0,0)$.
But now, the whole circle is allowed.

\begin{figure}[h!]
\centering
\includegraphics[width=0.49\linewidth]{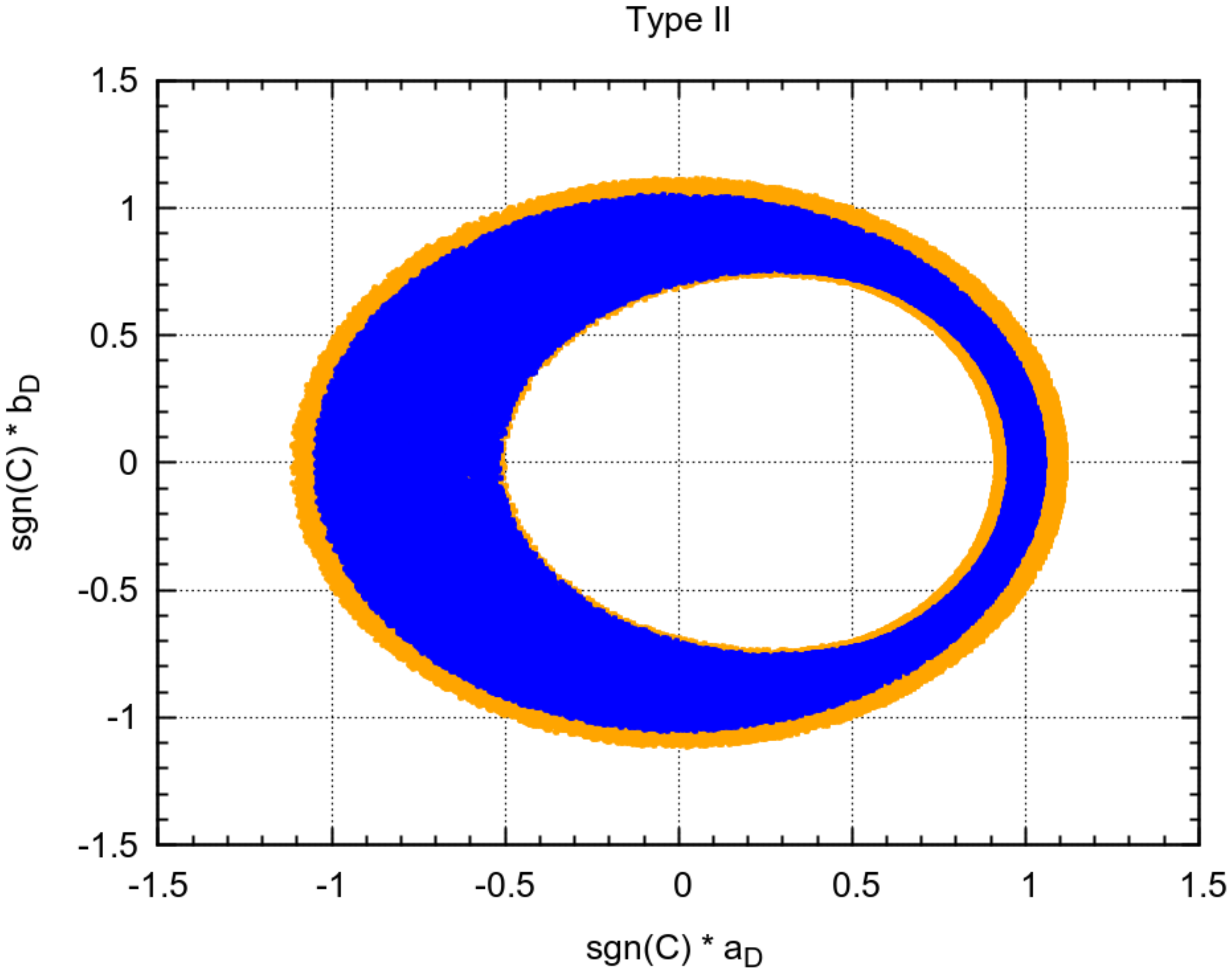}
\hspace{-0.02\linewidth}
\includegraphics[width=0.49\linewidth]{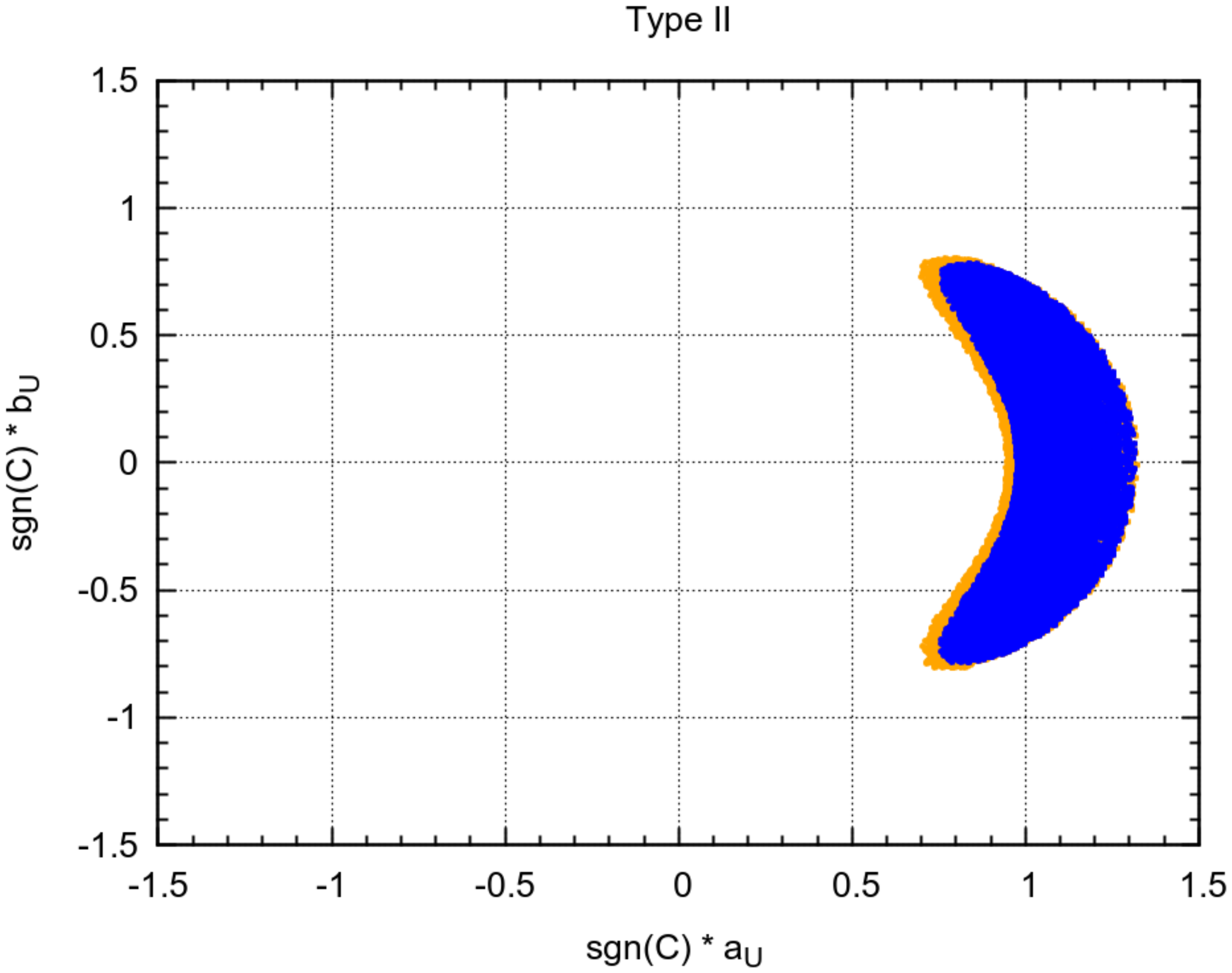}
\caption{Left: sgn$(C)$ $b_D$ $=$ sgn$(C)$ $b_L$ as a function of sgn$(C)$ $a_D$ $=$ sgn$(C)$ $a_L$  
for Type II and a center of mass
energy of $13$ TeV with $\mu_{VV}$ within $20\%$ (orange/light-grey)
and $10\%$ (blue/black) of the SM.
}
\label{fig:JR1}
\end{figure}

The constraints from $\mu_{VV}$ can be understood with
simple arguments as follows.
It was shown in \cite{Fontes:2014tga},
in the real 2HDM, that the limits on $\mu_{VV}$ impose rather
non trivial constraints on the coupling to fermions which,
however, can be understood from simple trigonometry.
Following the spirit of that article,
we assume that the production is mainly due to
$gg \rightarrow h_1$ with an intermediate top in the triangle loop,
and that the scalar decay width is dominated by the
decay into $b \bar{b}$.
As a result,
\be
\mu_{VV} \sim
(a_U^2 + 1.5\, b_U^2)\, \frac{C^2}{a_D^2 + b_D^2},
\ee
where the approximate factor of 1.5 is what one would obtain
either from a naive one-loop calculation\footnote{See,
for instances, equations (A7)-(A9) and (A14) in \cite{Barroso:2012wz}.},
or from a full HIGLU simulation \cite{Spira:1995mt}.
Applying this formula, 
we obtain figure~\ref{fig:JR1},
where we have taken $\mu_{VV}$ within $20\%$ (orange/light-grey)
or $10\%$ (blue/black) of the SM,
letting the angles vary freely within their theoretically allowed ranges.
%
%
The similarity between the left (right) panes
of figures~\ref{fig:NEW} 
(a full model simulation) and figure~\ref{fig:JR1}
(a simple trigonometric exercise) is uncanny.

Further constraints are brought about by a second
simple geometrical argument. They place all solutions close to
$(a,b) \sim (1,0)$ when $C$ is close to unity.
We use eqs.~\eqref{RsUP} to derive
\begin{equation}
1=R_{11}^2 + R_{12}^2 + R_{13}^2=
\frac{(C - s_\beta^2 \, a_U)^2}{c_\beta^2}
+ s_\beta^2 \, a_U^2 + \tan^2{\beta} \, b_U^2,
\end{equation}
leading to
\begin{equation}
\left( a_U - C \right)^2 + b_U^2
=
\frac{1}{\tan^2{\beta}}\, [1-C^2].
\label{circleUP}
\end{equation}
This is a circle centered at $(C,0)$,
which excludes most cyan/light-grey points on
the right panel of figure~\ref{fig:NEW}.
Since $C$ is close to unity,
and appears divided by $\tan{\beta}$ (which must be larger than one),
the radius is almost zero,
forcing $a_U$ to lie close to $C \sim 1$,
and $b_U$ close to $0$.
Including all channels at 10\%
restricts the region even further,
as seen in the blue/black points
on the right panel of figure~\ref{fig:NEW}.
It is true that an equation similar to
eq.~\eqref{circleUP} can be found for the down sector:
\begin{equation}
\left( a_D - C \right)^2 + b_D^2
=
\tan^2{\beta}\, [1-C^2].
\label{circleDOWN}
\end{equation}
However,
the different placement of $\tan{\beta}$ is crucial.
For intermediate to large $\tan{\beta}$,
the $\tan^2{\beta}$ factor in eq.~\eqref{circleDOWN}
enhances the radius with respect to that
allowed by the $\cot^2{\beta}$ factor in
eq.~\eqref{circleUP}.
This explains the difference between the red/dark-grey points on the
two panels in figure~\ref{fig:F4}.

We now turn to the constraints on the $\sin \alpha_2$-$\tan{\beta}$ plane.
%
\begin{figure}[h!]
\centering
\includegraphics[width=0.49\linewidth]{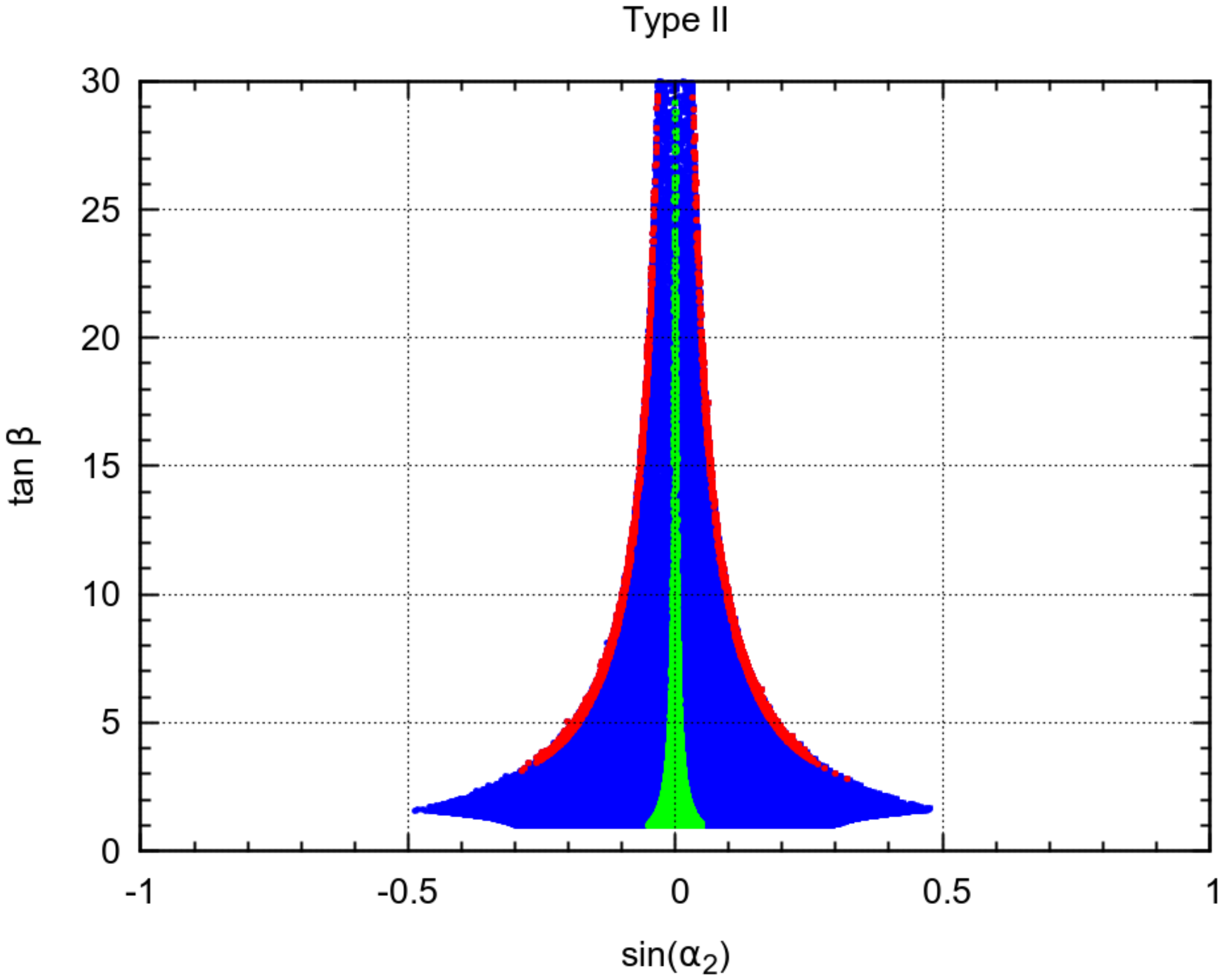}
\hspace{-0.02\linewidth}
\includegraphics[width=0.49\linewidth]{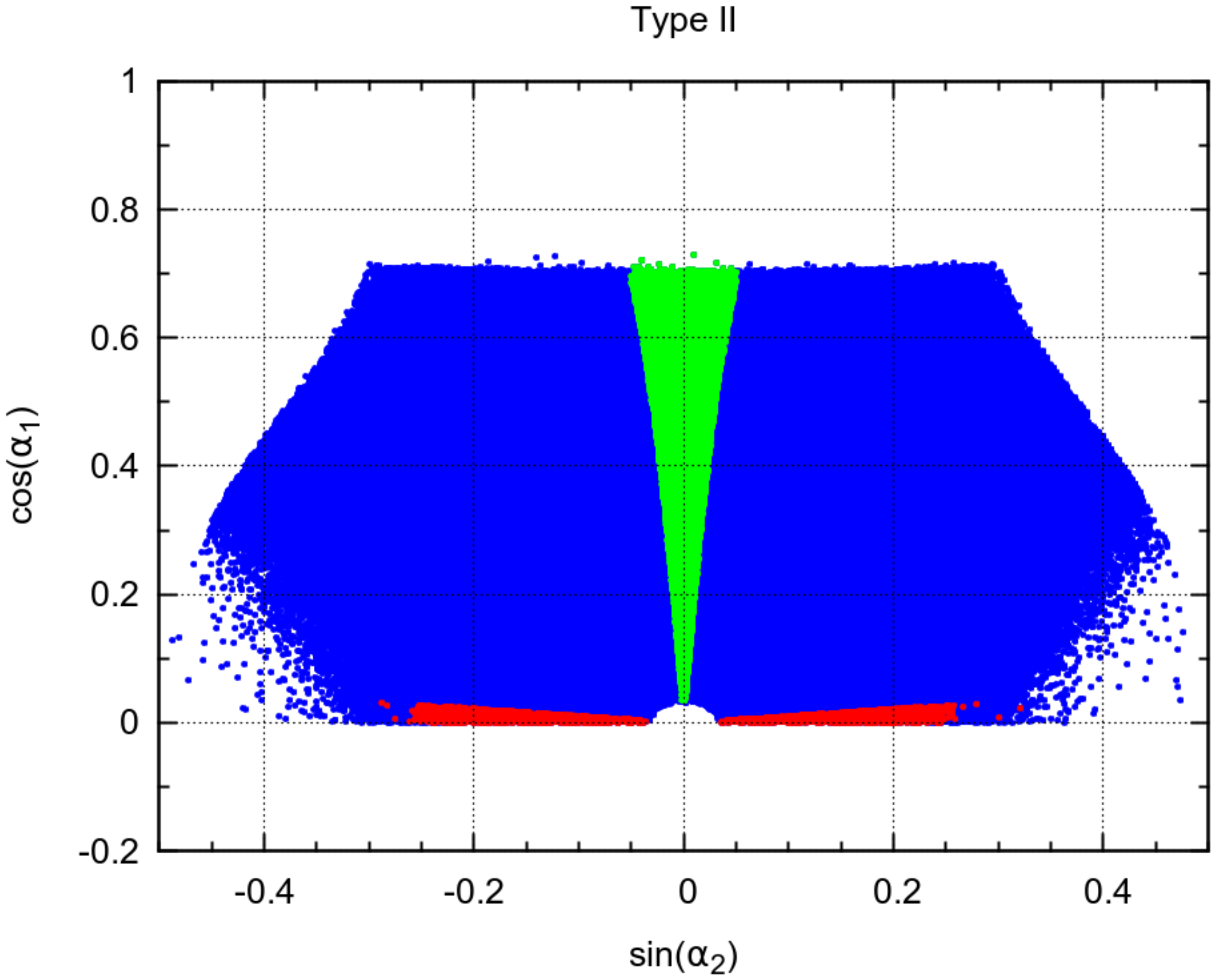}
\caption{Left: $\tan \beta$ as a function of $\sin \alpha_2$ for Type II and a center of mass
energy of $13$ TeV with all rates
at $10$\% (blue/black). In red/dark-grey we show the points with $|a_D| < 0.1$ and $||b_D|-1| < 0.1$ and
in green $|b_D| < 0.05$ and $||a_D|-1| < 0.05$
Right: same, with $\tan \beta$ replaced by $\cos \alpha_1$.}
\label{fig:F4b}
\end{figure}
%
When we choose $\mu_{VV} > 0.9$ in the exact limit $(|a_D|,|b_D|)=(0,1)$,
we obtain, using the approximation in eq. (\ref{approxtype2})
$\tan \beta > 4.4$. Because we are not in the exact limit, the bound we present in the left
plot of figure~\ref{fig:F4b} for $\tan \beta$ is closer to $3$. The left panel of
figure~\ref{fig:F4b} shows $\tan \beta$ as a function of $\sin \alpha_2$ for Type II
and a center of mass energy of $13$ TeV with all rates at $10$\% (blue/black). In red/dark-grey we show
the points with $|a_D| < 0.1$ and $||b_D| - 1| < 0.1$ and
in green $|b_D| < 0.05$ and $||a_D| - 1| < 0.05$. In the right panel, $\tan \beta$
is replaced by $\cos \alpha_1$. These two plots allow us to distinguish the main features
of the SM-like scenario, where $(|a_D|,|b_D|) \approx (1,0)$ from the pseudoscalar
scenario where $(|a_D|,|b_D|) \approx (0,1)$. In the SM-like scenario $\sin \alpha_2 \approx 0$,
$\tan \beta$ is not constrained and the allowed values of $\sin \alpha_2$ grow with increasing
$\cos \alpha_1$. In the pseudoscalar scenario $\cos \alpha_1 \approx 0$, $\sin \alpha_2$
and $\tan \beta$ are strongly correlated and $\tan \beta$ has to be above $\approx 3$.
Clearly, all values of $a_D$ and $b_D$ are allowed provided $a_D^2 + b_D^2 \approx 1$.

\begin{figure}[h!]
\centering
\includegraphics[width=0.49\linewidth]{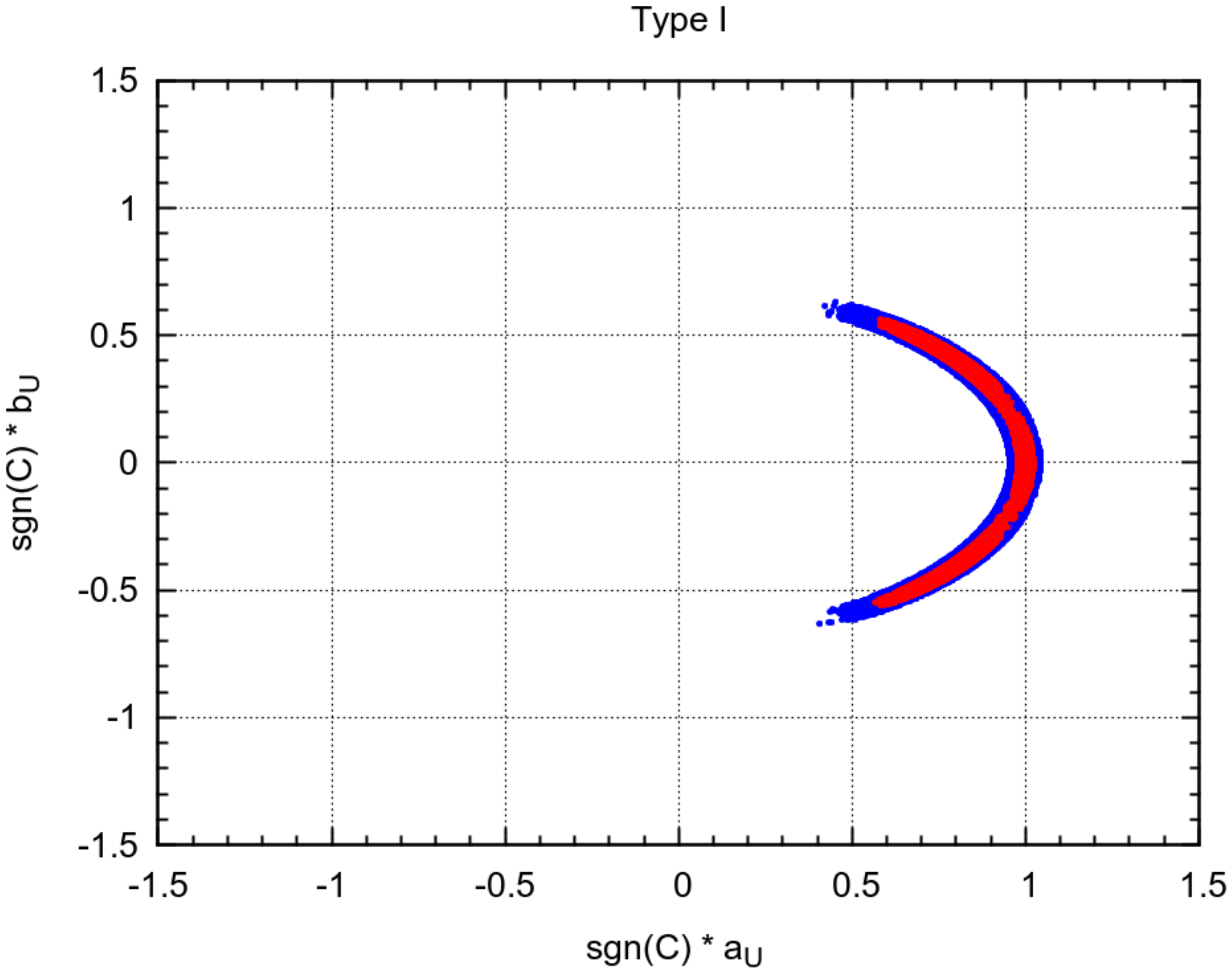}
\hspace{-0.02\linewidth}
\includegraphics[width=0.49\linewidth]{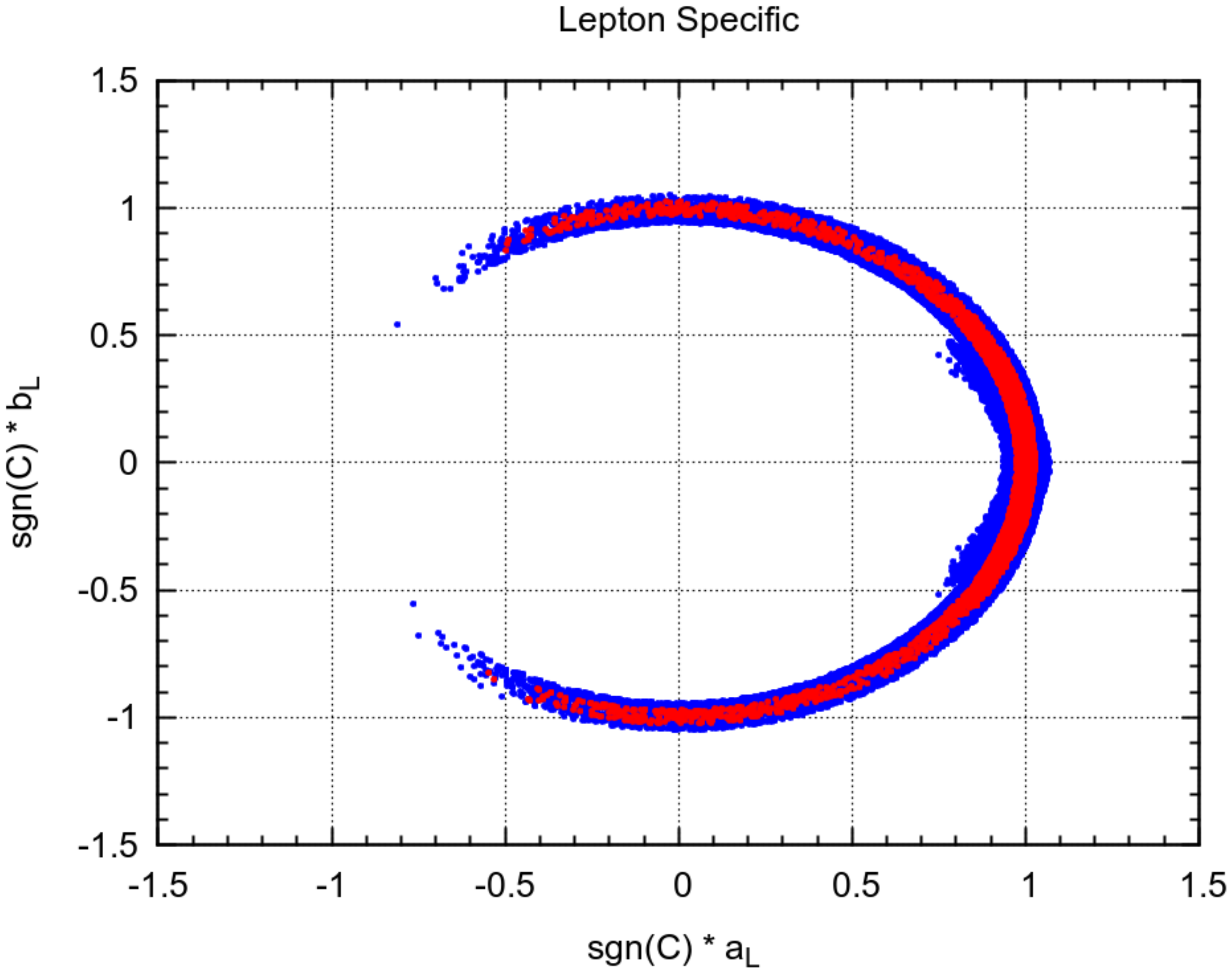}
\caption{Left: sgn$(C)$ $b_U$ as a function of sgn$(C)$ $a_U$ for Type I and a center of mass
energy of $13$ TeV with all rates
at $10$\% (blue/black) and $5$\% (red/dark-grey).
Right: sgn$(C)$ $b_L$ as a function of sgn$(C)$ $a_L$ for LS and a center of mass
energy of $13$ TeV with all rates
at $10$\% (blue/black) and $5$\% (red/dark-grey).}
\label{fig:F5}
\end{figure}
%
In the left panel of figure~\ref{fig:F5} we show $b_U$ as a function of
$a_U$ for Type I and a center of mass energy of $13$ TeV with all rates
at $10$\% (blue/black) and $5$\% (red/dark-grey). In Type I this plot is valid for all
Yukawa couplings, because $a_U=a_D=a_L$ and $b_U=b_D=b_L$.
It is interesting that even at $10$\%
there are points close to $(a,b)=(0.5,0.6)$ still allowed
and no dramatic changes happen when we move to $5$\%. In the right
plot we show $b_L$ as a function of $a_L$ for LS  with the same colour code.
Here again the $(a_L,b_L)=(0,1)$ scenario is still
allowed both with $10$\% and $5$\% accuracy. However, as was previously shown,
the wrong sign limit is not allowed for the LS model~\cite{Fontes:2014tga, Ferreira:2014dya}.
Nevertheless, in the C2HDM, the scalar component sgn$(C) \, a_L$
can  reach values close to $-0.8$. Finally, for the up-type and down-type quarks, the plots are very similar
to the one in the right panel of figure~\ref{fig:F4} for Type II.

\subsection{Direct measurements of the CP-violating angle}

Although precision measurements already constrain both the scalar
and pseudoscalar components of the Yukawa couplings in the C2HDM,
there is always the need for a direct (and thus, more model independent)
measurement of the relative size of pseudoscalar to
scalar components of the Yukawa couplings.
The angle that measures
this relative strength, $\phi_i$, defined as
\begin{equation}
\tan \phi_i  = b_i/a_i \qquad i=U,\, D, \, L \, ,
\end{equation}
could in principle be measured for all Yukawa couplings. The experimental
collaborations at CERN will certainly tackle this problem when the
high luminosity stage is reached, through
any variables able to measure the ratio of the pseudoscalar to scalar
component of the Yukawa couplings. There are several proposals for a direct measurement
of this ratio, which focus mainly on the $tth$ and on the $\tau^+ \tau^- h$ couplings.
Measurement of $b_U/a_U$ were first proposed for $pp \to t \bar t h$
in~\cite{Gunion:1996xu} and more recently reviewed in~\cite{Ellis:2013yxa, He:2014xla, Boudjema:2015nda}.
A proposal to probe the same vertex through the process $pp \to hjj$~\cite{DelDuca:2001fn} was
put forward in~\cite{Field:2002gt} and again more recently in~\cite{Dolan:2014upa}.
In reference~\cite{Dolan:2014upa} an exclusion of $\phi_t > 40 \degree$ ($\phi_t > 25 \degree$)
for a luminosity of 50 fb$^{-1}$ (300 fb$^{-1}$) was obtained for 14 TeV
and assuming $\phi_t=0$ as the null hypothesis. A study of the $\tau^+ \tau^- h$
vertex was proposed in~\cite{Berge:2008wi} and a detailed study taking into account the
main backgrounds~\cite{Berge:2014sra} lead to an estimate in the precision
of $\Delta \phi_\tau$ of $27 \degree$ ($14.3 \degree$) for a luminosity of 150 fb$^{-1}$
(500 fb$^{-1}$) and a center of mass energy of 14 Tev.

Since in the C2HDM the couplings are not universal, one would need in principle
three independents measurements, one for up-type quarks, one for down-type quarks and one for leptons.
The number of independent measurements is of course model dependent. For Type I one
such measurement is enough because the Yukawa couplings are universal. For all other
Yukawa types we need two independent measurements. It is interesting to note that
for model F, since the leptons and up-type quarks coupling to the Higgs are the same,
a direct measurement of the $hbb$ vertex is needed to probe the model. On the other
hand, and again using model F as an example, a different result for $\phi_t$ and
$\phi_\tau$ would exclude model F (and also Type I).

\begin{figure}[h!]
\centering
\includegraphics[width=0.32\linewidth]{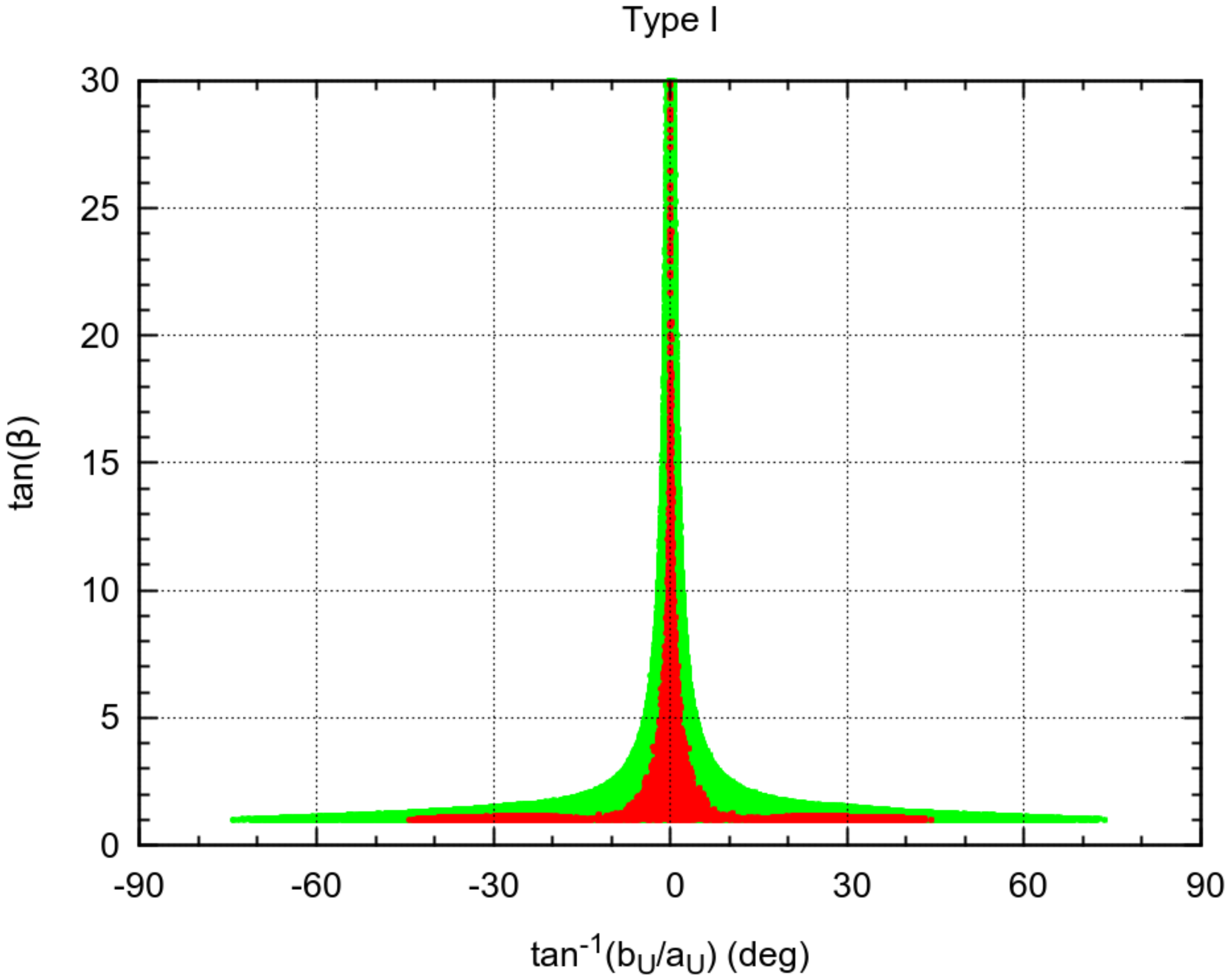}
\hspace{-0.02\linewidth}
\includegraphics[width=0.32\linewidth]{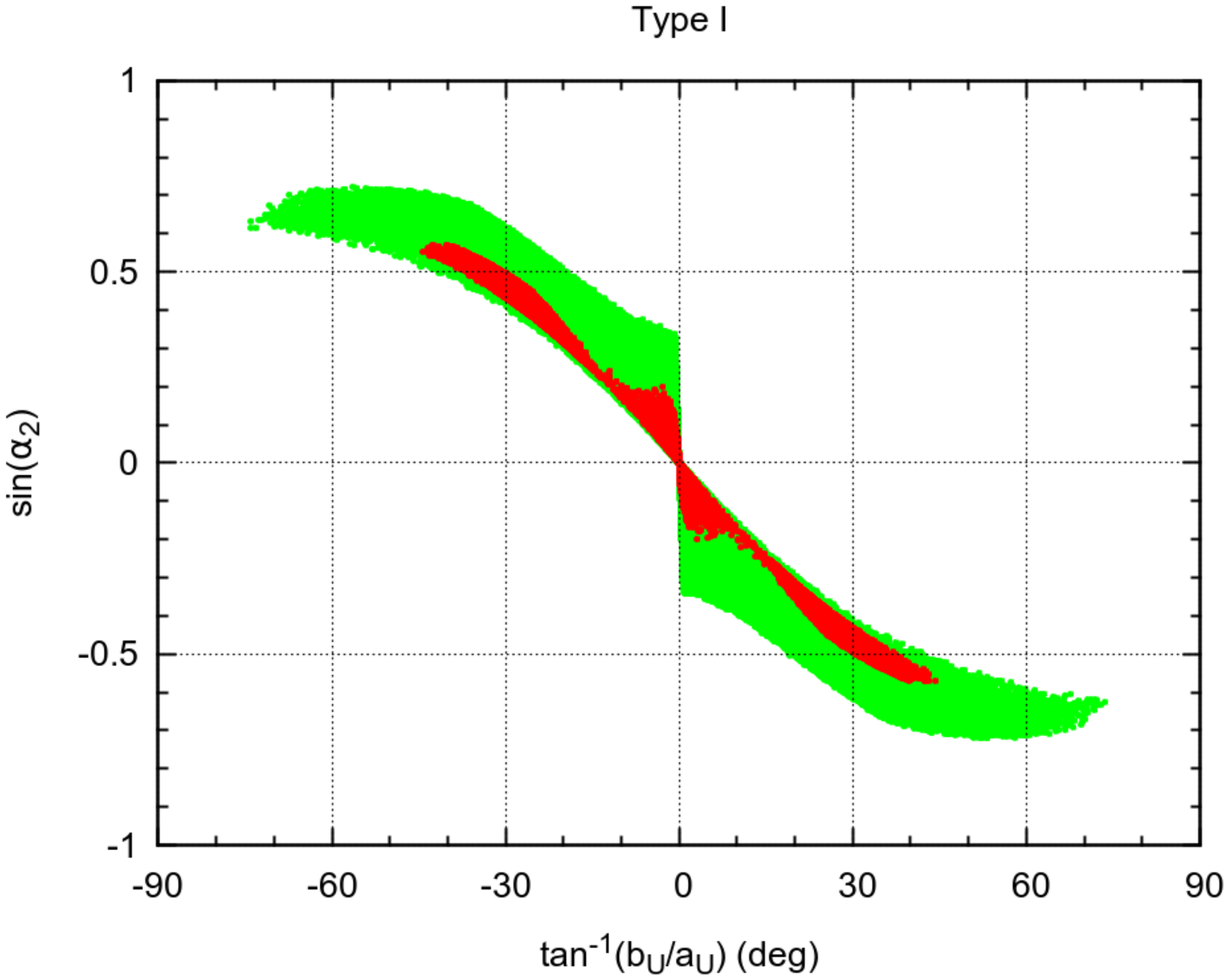}
\hspace{-0.02\linewidth}
\includegraphics[width=0.32\linewidth]{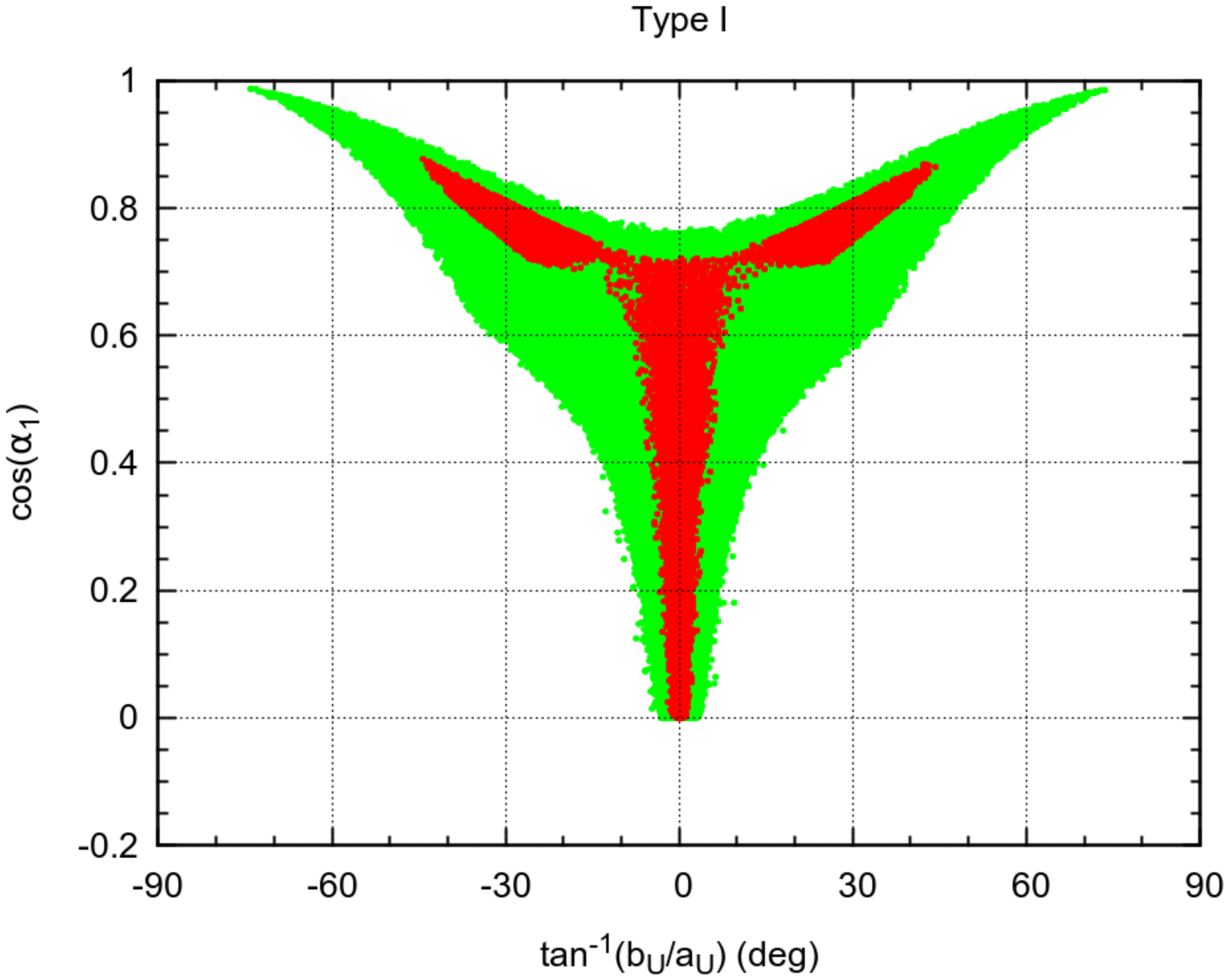}
\vspace{1mm}\\
\includegraphics[width=0.32\linewidth]{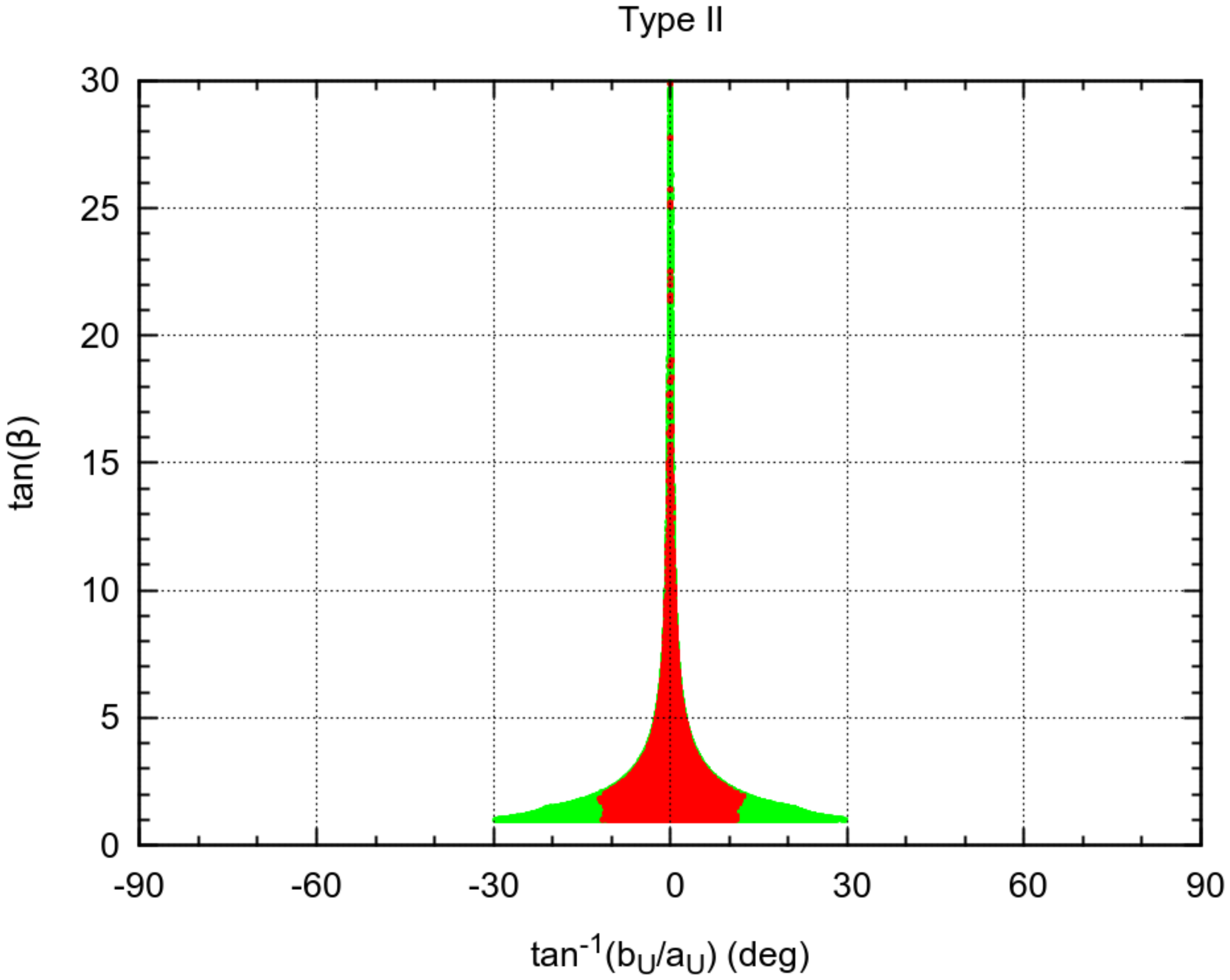}
\hspace{-0.02\linewidth}
\includegraphics[width=0.32\linewidth]{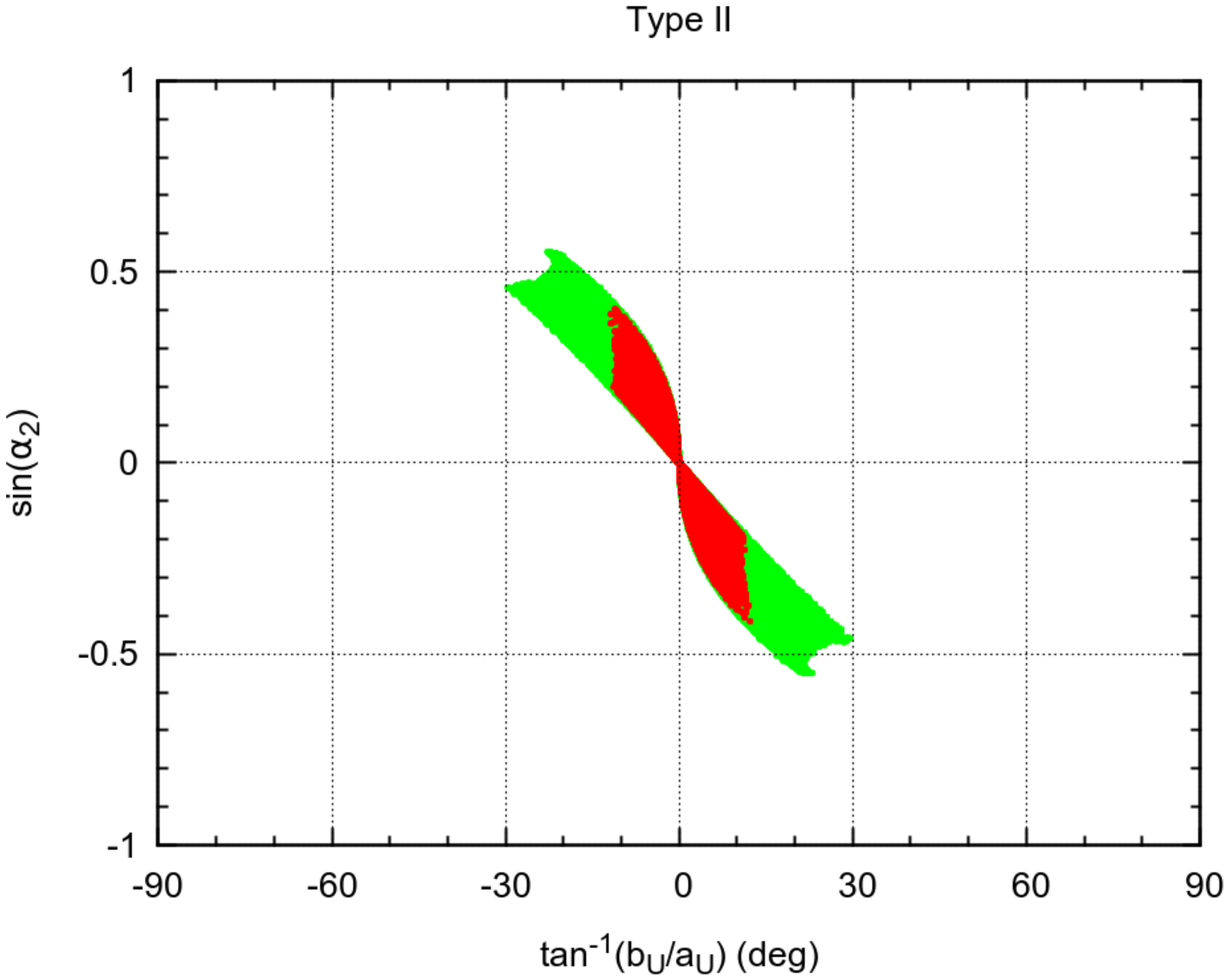}
\hspace{-0.02\linewidth}
\includegraphics[width=0.32\linewidth]{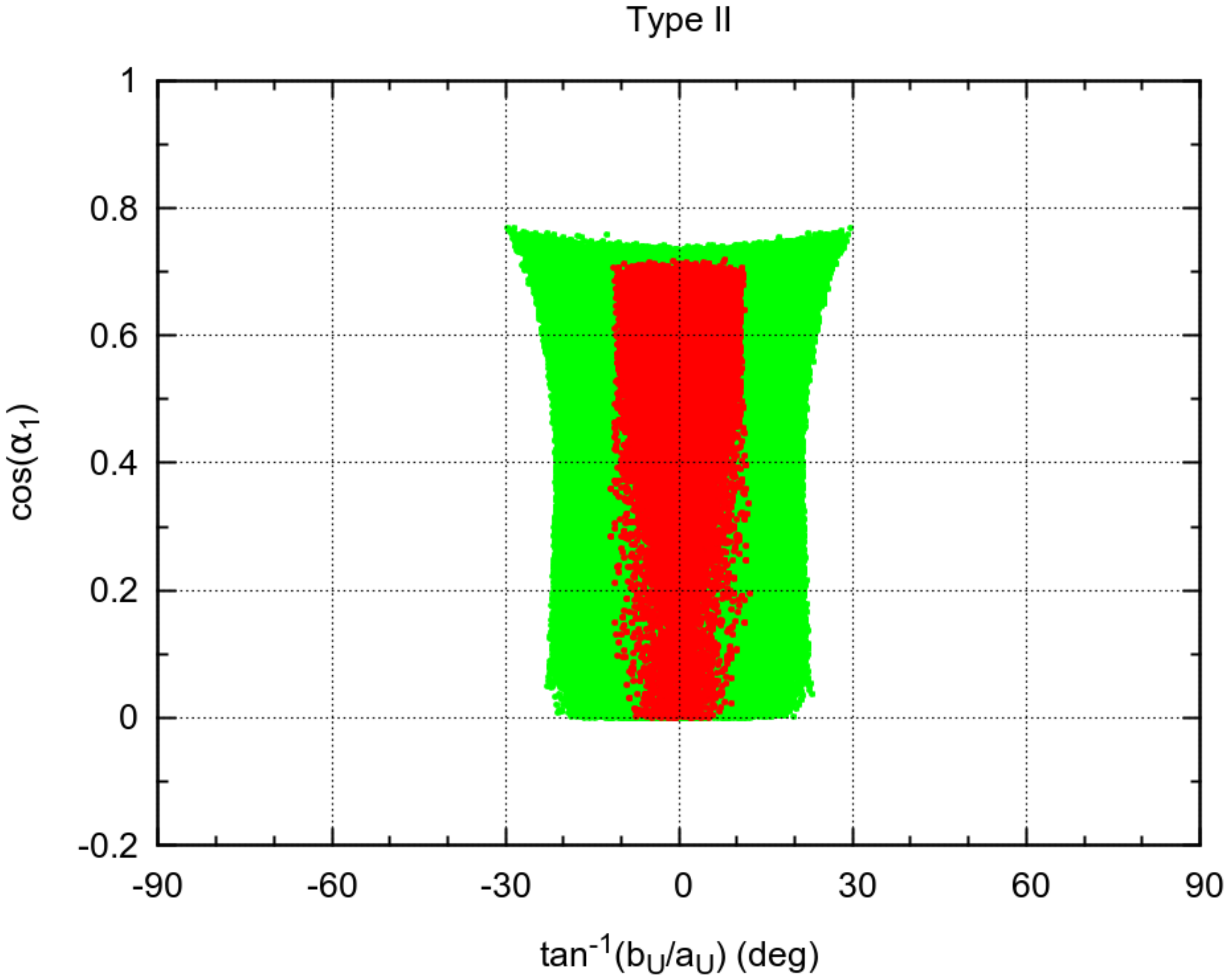}
\caption{Top: $\tan \beta$ (left), $\sin \alpha_2$ (middle) and  $\cos \alpha_1$ (right)
as a function of $\phi_U=\tan^{-1} (b_U/a_U)$ for Type I
with rates at $20$\% (green) and $5$\% (red/dark-grey). Bottom: same but for Type II.}
\label{fig:F6}
\end{figure}
Let us first discuss what we can already say about the allowed range for the $\phi_U \equiv \phi_t$
angle and what to expect by the end of the LHC's run 2 using only
the rates' measurements. In figure~\ref{fig:F6} we show on the top row $\tan \beta$ (left), $\sin \alpha_2$ (middle)
and  $\cos \alpha_1$ (right) as a function of $\phi_U=\tan^{-1} (b_U/a_U)$ for Type I,
with rates within $20$\% (green) and $5$\% (red/dark-grey) of the SM prediction. In the bottom row we present the same
plots but for Type II. The green points are a good approximation for the allowed region after run 1,
while the red/dark-grey points are a good prediction for the allowed space with the run 2
high luminosity results. The most striking features of the plots are the following. For Type
I the angle $\phi_U=\phi_D=\phi_L$ is between $-75 \degree$ and $75 \degree$ and this interval
will be reduced to roughly $-45 \degree$ and $45 \degree$ provided the measured rates are in agreement
with the SM predictions. For Type II only $\phi_U$ is constrained; we get $|\phi_U| < 30 \degree$
and the prediction of roughly $|\phi_U| < 15 \degree$ when rates are within 5$\%$ of the SM predictions.
Since the Higgs couplings to top
quarks are the same for all models, the angle that relates scalar and pseudoscalar
components for this vertex is related to the lightest Higgs CP-violating angle $\alpha_2$ by
\begin{equation}
\tan \phi_t  = - c_\beta/s_1 \, \tan \alpha_2 \qquad  \Rightarrow \qquad \tan \alpha_2 = - s_1/c_\beta \, \tan \phi_t \, .
\label{eq:phi1}
\end{equation}
The parameter space is restricted in such a way that high $\tan \beta$ implies low $\alpha_2$. Since
$s_1$ cannot be too small, it is clear from equation~(\ref{eq:phi1}) that large $\tan \beta$ necessarily
implies a small $\phi_t$. This is clearly seen in the left top and bottom plots of figure~\ref{fig:F6}
where for large $\tan \beta$ the pseudoscalar component of the up-type quarks Yukawa coupling is very close to
zero. Interestingly, for both Type I and Type II the values of $\tan \beta \thicksim$ O$(1)$ are the ones for which
the angle $\phi_t$ is less constrained. These are exactly the values for which the the coupling
$tth$ has a maximum value (already considering the remaining constraints that disallow values of $\tan \beta$
below 1). Therefore, a direct measurement of $\phi_t$ could still be competitive with the rates measurement
in Type I.

%
\begin{figure}[h!]
\centering
\includegraphics[width=0.49\linewidth]{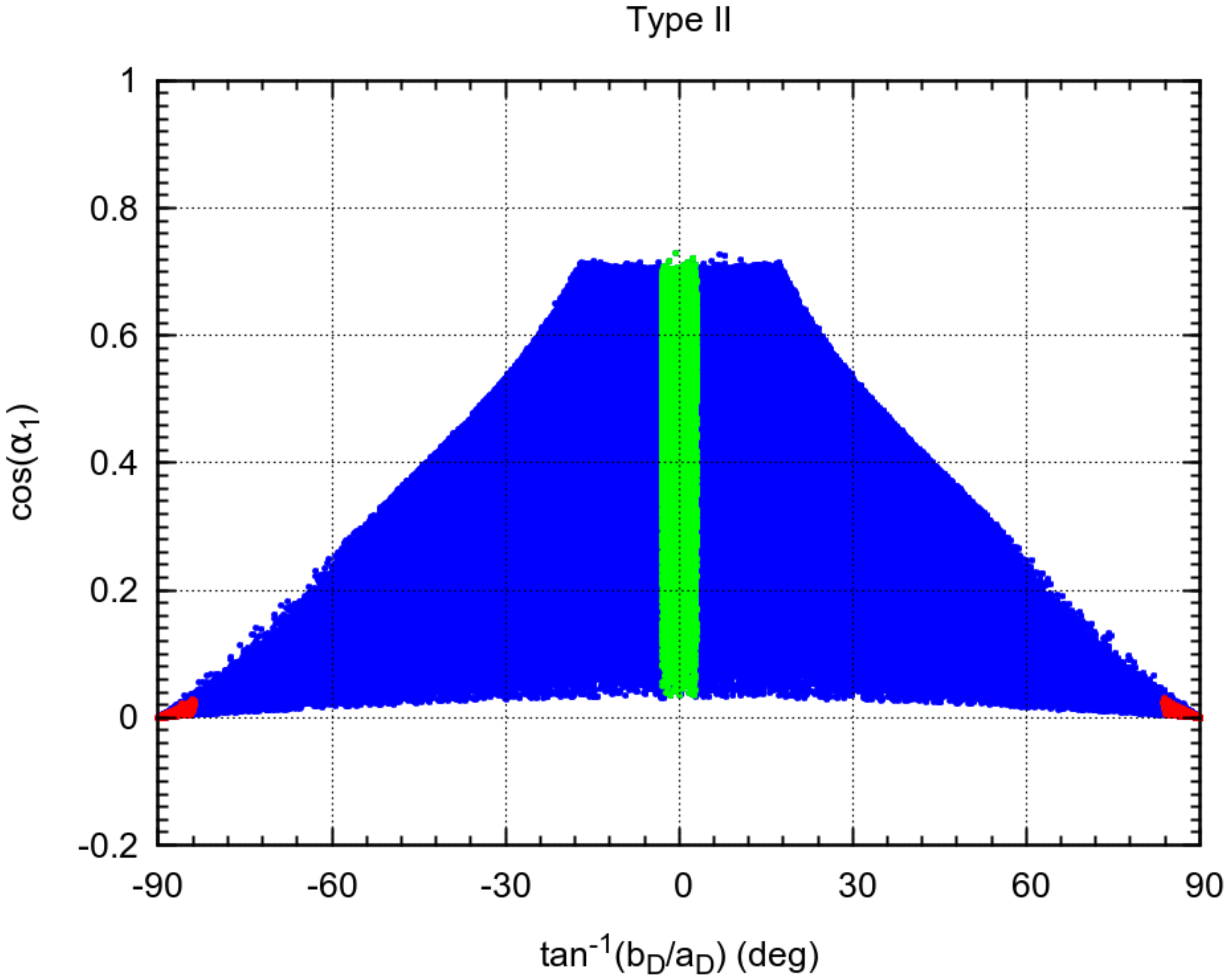}
\hspace{-0.02\linewidth}
\includegraphics[width=0.49\linewidth]{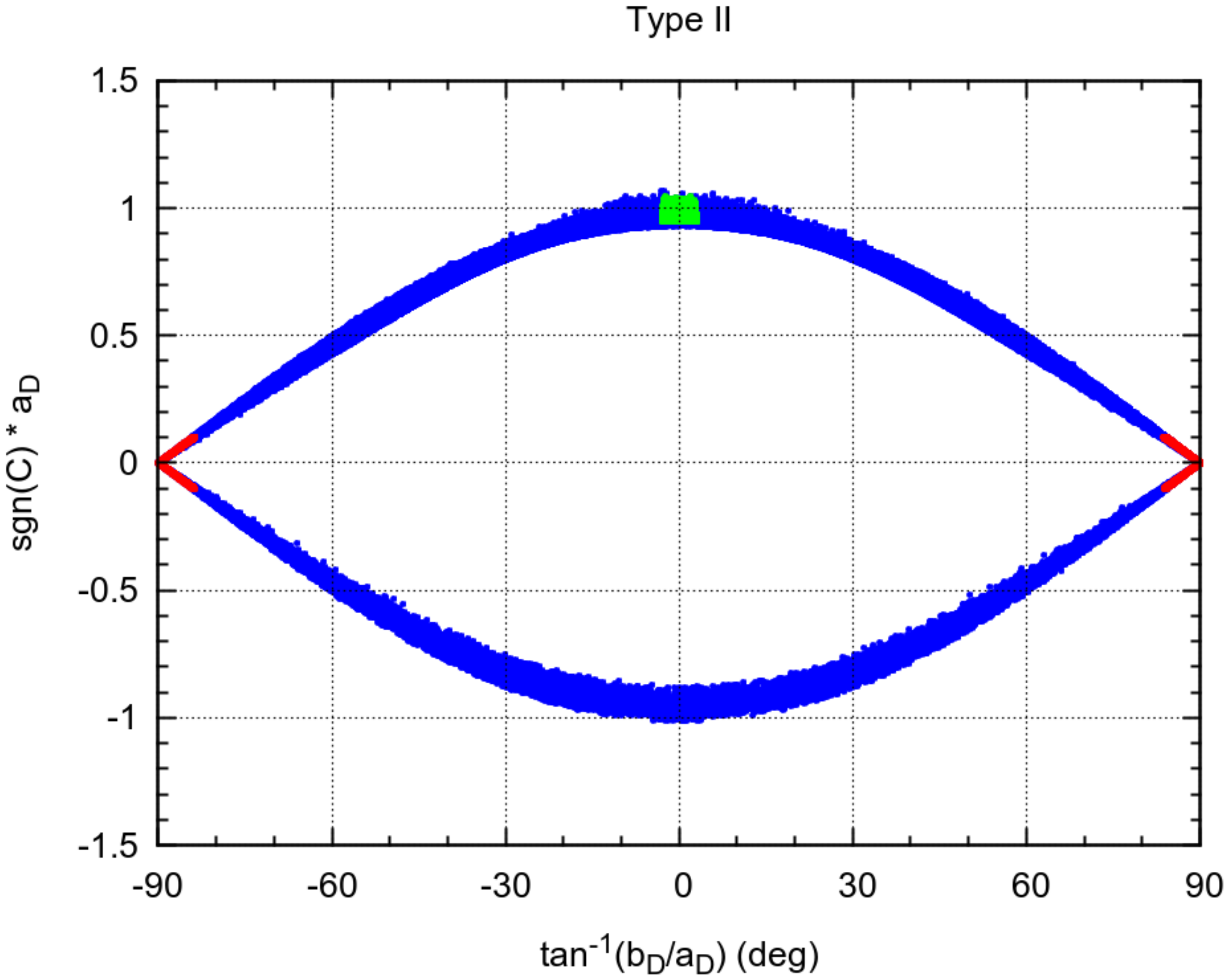}
\caption{Left: $\cos \alpha_1$ as a function of $\tan^{-1} (b_D/a_D)$ for Type II and a center of mass
energy of $13$ TeV with all rates
at $10$\% (blue/black). In red/dark-grey we show the points with $|a_D| < 0.1$ and $||b_D|-1| < 0.1$ and
in green $|b_D| < 0.05$ and $||a_D|-1| < 0.05$.
Right: same, with $\cos \alpha_1$ replaced by sgn$(C) \, a_D$.}
\label{fig:F7}
\end{figure}
%
Let us now move to the Yukawa versions that can have a zero scalar component not only at the end of run 1,
but also at the end of run 2, if only the rates are considered. For definiteness we focus on Type II.
As previously discussed, a direct measurement involving the vertex
$h \tau^+ \tau^-$~\cite{Berge:2008wi, Berge:2014sra} could lead to a precision
in the measurement of $\phi_\tau$, $\Delta \phi_\tau$, of $27 \degree$ ($14.3 \degree$) for a luminosity of 150 fb$^{-1}$
(500 fb$^{-1}$) and a center of mass energy of 14 Tev. In figure~\ref{fig:F7}
(left) we show $\cos \alpha_1$ as a function of $\tan^{-1} (b_D/a_D)$ for Type II and a center of mass
energy of $13$ TeV with all rates
at $10$\% (blue/black). In red/dark-grey we show the points with $|a_D| < 0.1$ and $||b_D|-1| < 0.1$ and
in green $|b_D| < 0.05$ and $||a_D|-1| < 0.05$. In the right panel $\cos \alpha_1$ is replaced by sgn$(C) \, a_D$.
It is clear that the SM-like scenario sgn$(C)$ $(a_D, \, b_D ) = (1,0)$ is easily distinguishable
from the $(0,1)$ scenario. In fact, a measurement of $\phi_\tau$ even if not very precise would easily
exclude one of the scenarios. Obviously, all other scenarios in between these two will need
more precision (and other measurements) to find the values of scalar and pseudoscalar components.
The $\tau^+ \tau^-  h$ angle is related to $\alpha_2$ as
\begin{equation}
 \tan \phi_\tau  = - s_\beta/c_1 \, \tan \alpha_2 \qquad  \Rightarrow \qquad
 \tan \alpha_2 = - c_1/s_\beta \, \tan \phi_\tau
\end{equation}
and therefore a measurement of the angle $\phi_\tau$ does not directly constrain
the angle $\alpha_2$. In fact, the measurement gives a relation between the three angles.
A measurement of $\phi_t$ and $\phi_\tau$ would give us two independent relations
to determine the three angles.


\subsection{Constraints from EDM}

Models with CP violation are constrained by bounds on the electric
dipole moments (EDMs) of neutrons, atoms and molecules.
Recently the ACME Collaboration~\cite{Baron:2013eja}
improved the bounds on the electron EDM by looking at the EDM
of the ThO molecule.
This prompted several groups to look again at the subject.
For what concerns us here, the complex 2HDM,
several analyses have been performed recently~\cite{Buras:2010zm,
Cline:2011mm, Jung:2013hka, Shu:2013uua, Inoue:2014nva, Brod:2013cka}.
In ref.~\cite{Inoue:2014nva} it was found that the most stringent limits
are obtained from the ThO experiment, except in cases where there are
cancellations among the neutral scalars.
These cancellations were pointed out in \cite{Jung:2013hka, Shu:2013uua}
and arise due to orthogonality of the $R$ matrix in the case of almost
degenerate scalars~\cite{Fontes:2014xva}.
So far, there is no complete scan of EDM in the C2HDM; only
some benchmark points have been considered, making it difficult to see
when these cancellations are present.

What can be learned from these studies is that the EDMs are very
important and their effect in the C2HDM has to be taken in account in
a systematic way, in the sense that, for each point in the scan, the
EDMs have to be calculated and compared with the experimental
bounds.
However, for the purpose of the studies in this work and for
the present experimental sensitivity, this is not required.
This is because we are looking at scenarios where the couplings
of the up-type sector (top quark) are very close to the SM and
the differences, still allowed by the LHC data, are in the couplings
of the down-type sector; the tau lepton and bottom quark.
As was shown in ref.~\cite{Brod:2013cka},
while the pseudoscalar coupling of the top quark is very much constrained
(in our notation $|b_U| \leq 0.01$),
the corresponding couplings for the b quark and tau lepton are less
constrained by the EDMs than by the LHC data.
Since we are taking in account the collider data,
our scenarios are in agreement with the present experimental data.
But one should keep in mind that, as pointed out in
refs.~\cite{Brod:2013cka, Dekens:2014jka},
the future bounds from the EDMs can alter this situation.
In the future, the interplay between the EDM bounds and the data from
the LHC
Run 2 will pose relevant new constraints in the complex 2HDM in
general, and in particular for the scenarios presented in this work.

\section{Conclusions}
\label{sec:conc}

We discuss the present status of the allowed parameter space of the
complex two-Higgs doublet model where we have considered all pre-LHC
plus the theoretical constraints on the model. We have also
taken into account the bounds arising from assuming that the lightest
scalar of the model is 125 GeV Higgs boson discovered at the LHC.
We have shown that the parameter space is already quite constrained
and recovered all the limits on the couplings of a CP-conserving 125 GeV Higgs.
The allowed space for some variables, as for example for the $\tan \beta$
parameter, is now increased as a natural consequence of having a larger
number of variables to fit the data as compared to the CP-conserving case.

The core of the work is the discussion of scenarios where the scalar component of the Yukawa
couplings of the lightest Higgs to down-type quarks and/or to leptons can vanish.
In these scenarios, that can occur for Type II, F and LS, the pseudoscalar component
plays the role of the scalar component in assuring the measured rates at the LHC.
A direct measurement of the angle that gauges the ratio of pseudoscalar to scalar
components in the $tth$ vertex, $\phi_t$, will probably help to further constrain
this ratio. However, it is the measurement of $\phi_\tau$, the angle for the
$\tau^+ \tau^- h$ vertex, that will allow to rule out the scenario of a vanishing scalar
even with a poor accuracy. We have also noted that for the F model only a direct
measurement of $\phi_D$ in a process involving the $bbh$ vertex would be able to probe
the vanishing scalar scenario. Finally a future linear collider~\cite{Ono:2012ah, Asner:2013psa} will certainly help
to further probe the vanishing scalar scenarios.

\begin{acknowledgments}
%
%
R.S. is supported in part by the Portuguese
\textit{Funda\c{c}\~{a}o para a Ci\^{e}ncia e Tecnologia} (FCT)
under contract PTDC/FIS/117951/2010.
D.F., J.C.R. and J.P.S. are also supported by the Portuguese Agency FCT
under contracts
CERN/FP/123580/2011,
EXPL/FIS-NUC/0460/\-2013 and PEst-OE/FIS/UI0777/2013.
\end{acknowledgments}

\end{document}